\documentclass[11pt,a4paper]{article}
\usepackage[english]{babel} 
\usepackage{amsmath}
\usepackage{mathtools}
\usepackage{amsfonts}
\usepackage{amssymb}
\usepackage{amsthm}
\usepackage{makeidx}
\usepackage{graphicx}
\usepackage{natbib}
\usepackage{setspace}
\usepackage{float}
\usepackage{hyperref}
\usepackage[margin=1in]{geometry}
\usepackage{setspace}
\usepackage[titletoc]{appendix}
\usepackage{cleveref}
\usepackage{systeme}
\usepackage{relsize}
\usepackage{multirow}
\usepackage{amsmath}
\usepackage{ragged2e}
\allowdisplaybreaks

\hypersetup{
	colorlinks   = true,
	citecolor    = blue,
	urlcolor = blue
}

\newtheorem{Def}{Definition}
 \newtheorem{Thm}{Theorem}
 \newtheorem{Prp}{Proposition}
 \newtheorem{Lmm}{Lemma}
 
 \theoremstyle{definition}
 \newtheorem{Exm}{Example}
 
 \theoremstyle{remark}
 \newtheorem{Rmk}{Remark}

\DeclareMathOperator*{\argmin}{arg\,min}
\DeclareMathOperator*{\essinf}{ess\,inf}
\DeclareMathOperator*{\esssup}{ess\,sup}

\DeclareMathOperator*{\clconv}{clconv}

\newcommand{\E}{{E}}

\title{Robust convex risk measures}
\author{Marcelo Brutti Righi$^{a}$\footnote{We thank Professor Marlon Moresco for insightful comments. We are grateful for the financial support of CNPq (Brazilian Research Council) projects number 302614/2021-4, 307779/2022-0, and 401720/2023-3  and Foundation for Research Support of the State of Rio Grande do Sul (FAPERGS) under project number 23/2551-0000901-3.}\\\small{\href{mailto:marcelo.righi@ufrgs.br}{marcelo.righi@ufrgs.br}} \and Fernanda Maria M\"{u}ller$^{a}$ \\ \small{\href{mailto:}{fernanda.muller@ufrgs.br}}}
\date{\small{$^{a}$\textit{Business School, Federal University of Rio Grande do Sul, Washington Luiz, 855, Porto Alegre, Brazil, zip 90010-460}}}

\begin{document}
\maketitle

\begin{abstract}
In this paper, we refine and generalize closed forms for worst-case law invariant convex risk measures with uncertainty sets based on: i) closed balls under $p$-norms and Wasserstein distance; and ii)  moment constraints involving mean and variance. We also characterize the argmax of the worst-case problem in both settings. From such general results, we illustrate our framework by developing explicit closed forms for  concrete examples of convex risk measures. 
 Furthermore, we use extensive numerical simulations in order to assess the impact of robustness on capital determination and portfolio optimization.

\end{abstract}

\section{Introduction}\label{sec:intro}

The theory of risk measures in mathematical finance has become mainstream, especially since the landmark paper of \cite{Artzner1999}. A risk measure is a functional defined on some set of random variables that represents the monetary value of the risk. 
Knightian uncertainty is critical to risk management because it limits perfect knowledge of distributions. In this context, decision-makers face the consequences of their risk assessments with partial knowledge of probabilities and random variables.

  For risk measures, to deal with such uncertainty, it is customary to consider a worst-case approach, which involves a risk measure that is a point-wise supremum of a base risk measure over a specified uncertainty set. A common approach is linked to scenarios, indicating that robustness is related to the chosen probability, as discussed in \cite{Wang2018}, \cite{Bellini2018}, and \cite{Fadina2023}, \cite{miao2025robust}, for instance. A more general possibility is to address uncertainty about the choice of the risk measure, as in \cite{Righi2019} and \cite{Wang2023}, for instance. In both cases, the uncertainty set remains fixed for any point in the domain; therefore, the analysis is well documented. For instance, the penalty term, a key feature in the literature of risk measures computed as the convex conjugate, is given as the lower semicontinuous convex envelope of the point-wise infimum of the individual penalty terms.

  A more intricate set-up involves uncertainty about the random variables and how they affect risk measures, which are closely linked to the uncertainty and risk of the model. 
  In this case, the uncertainty depends on the random variables, since there are uncertainty sets specific to any point in the domain. In this paper, we refine and generalize closed forms for worst-case law invariant convex risk measures uncertainty sets based on: i) closed balls under $p$-norms and Wasserstein distance; and ii) mean and variance.  We also obtain a characterization for the argmax of the worst-case problem in both cases.

 In \Cref{thm:wasserstein}, we show that worst-case law invariant convex risk measures over closed balls in the $p$-norm and the Wasserstein distance coincide, based on a penalized convex conjugate. These results complement the literature on specific cases and risk measures.  \cite{Liu2022} obtains an affine closed formulation for spectral/distortion risk measures. In \cite{Bartl2020} and \cite{Li2023}, the worst-case of optimized certainty equivalents and shortfall risks over such balls is investigated, and \cite{Hu2024} studies the case of expectiles. 
 
 In \Cref{prp:mean}, we show that worst-case law invariant convex risk measures over uncertainty sets based on mean and variance are given as a mean plus maximum of a combination of standard deviation and convex conjugate.  This representation is a generalization over the ones obtained for spectral/distortion risk measures in \cite{Li2018}, \cite{Cornilly2018A}, \cite{Cornilly2019}, \cite{Cai2023}, \cite{Pesenti2020A}, \cite{Chen2021}, \cite{Shao2023}, \cite{Shao2023b}, and \cite{Zuo2024}. 

Propositions \ref{prp:ent} to \ref{prp:ES} present concrete instances of law-invariant convex risk measures within our framework. Specifically, we extend existing theory by deriving closed-form solutions for the Entropic Risk Measure, Shortfall Risk, Optimized Certainty Equivalent (OCE), Mean-plus-Semi-Deviation, Expectile, and Expected Shortfall.  These results represent a contribution since such formulations nest others in the literature or are totally new, easing their implementation into practical matters.

To the best of our knowledge, this is the first study to address these features in the context of law-invariant convex risk measures. The goal of most papers in this stream (see the mentioned paper above) is to develop closed forms over specific uncertainty sets, mostly for distortion risk measures or other specific classes of risk measures. Other papers address uncertainty sets of risk measures in a more general setup, such as \cite{Moresco2023}, which examines a dynamic setting, the interplay between the properties of risk measures and uncertainty sets, and \cite{Righi2024}, which develops risk measures over sets of random variables. However, none of these contributions considers the features examined in the present study.

This characterization plays a pivotal role in robust optimization problems. Intuitively, this approach introduces an adversary whose problem is inner maximization to account for the impact of the model uncertainty. Such worst-case situations are naturally difficult to address in terms of optimization. In this sense, recent work has been considered, especially by showing the problem is equivalent to usual convex ones or even finite-dimensional as in \cite{Pflug2012}, \cite{Wozabal2014}, \cite{Cai2023},  \cite{Pesenti2020A}, \cite{Pesenti2023B}, \cite{Blanchet2022}, \cite{Li2018}, \cite{Chen2021}, \cite{Liu2022}. 

The theoretical results are complemented by numerical experiments based on simulated data, which illustrate the worst-case forecast and its implications for portfolio allocation. The analysis is divided into two parts: examining how different uncertainty sets affect worst-case risk forecasts, and then applying the developed worst-case measures to a portfolio optimization problem. We aim to illustrate the worst-case measures under different uncertainty sets that reflect characteristics observed in real-world financial settings. This empirical investigation contributes to the discussion raised by \citet{xidonas2020robust}, who point out that, despite the abundance and practical relevance of the literature on robust portfolio optimization, there remains a general lack of empirical studies evaluating how these methods behave in applied contexts. Moreover, as emphasized by the authors, while some studies compare robust strategies to competing approaches, comparisons across robust strategies employing different types of risk measures — particularly convex law-invariant ones — are still relatively rare. Our experimental study addresses this gap by contrasting how distinct uncertainty structures (e.g., closed balls under $p$-norms and Wasserstein distance vs. moment constraints) influence both risk forecasts and optimal portfolio composition. These results are compared to the counterpart that is not based on uncertainty sets, i.e., the non-robust risk measure, as we refer to it.



\section{Background}\label{main}

 Consider the real-valued random result $X$ of any asset ($X\geq0$ is a gain and $X<0$ is a loss) defined on an atomless probability space $(\Omega,\mathcal{F},\mathbb{P})$. All equalities and inequalities are considered almost surely in $\mathbb{P}$. We define $X^+=\max(X,0)$, $X^-=\max(-X,0)$, and $1_A$ as the indicator function for an event  $A$. Let $L^{p}:=L^{p}(\Omega,\mathcal{F},\mathbb{P})$ be the space of (equivalent classes of) random variables such that  $ \lVert X \rVert_p^p = E[|X|^p]<\infty$ for $p\in[1,\infty)$ and $ \lVert X \rVert_\infty = \esssup |X| < \infty$ for $ p = \infty$, where $E$ is the expectation operator. Further, let $F_{X}(x) = P(X\leq x)$ and $F_{X}^{-1}(\alpha)=\inf\{x\in\mathbb{R}\colon F_X (x)\geq\alpha\}$ for $\alpha \in (0,1)$ be, respectively, the distribution function and the (left) quantile of $X$. For any $A\subseteq L^p$, we define $\mathbb{I}_A$ as its characteristic function on $L^p$, which assumes $0$ if $X\in A$, and $\infty$, otherwise. 
 
 When not explicit, it means that definitions and claims are valid for any fixed $L^p,\:p\in[1,\infty]$ with its usual p-norm. We denote by $\clconv$ the closed convex hull of a set in $L^p$. As usual, $L^q$, $\frac{1}{p}+\frac{1}{q}=1$ is the usual dual of $L^p$.  For $L^\infty$, we consider the dual pair $(L^\infty,L^1)$, where we call the weak topology its weak* topology. Let $\mathcal{Q}$ be the set of all probability measures on $(\Omega,\mathcal{F})$ that are absolutely continuous with respect to $\mathbb{P}$, with Radon--Nikodym derivative $\frac{d\mathbb{Q}}{d\mathbb{P}}\in L^q$. With some abuse of notation, we treat probability measures as elements of $L^q$. We denote $E_\mathbb{Q}$ for the expectation over $\mathbb{Q}$. For any $f\colon L^p\to\mathbb{R}$, its sub-gradient at $X\in L^p$ is $\partial f(X)=\{Q\in L^q\colon \rho(Z)-\rho(X) \geq E[(Z-X)Q]\:\forall\:Z\in L^p\}$.  We say $f\colon L^p\to\mathbb{R}$ is Gâteaux differentiable at $X\in L^p$ when $t\mapsto\rho(X+tZ)$ is differentiable at
$t = 0$ for any $Z\in L^p$ and the derivative defines a continuous linear functional on $L^p$. In the context of the Gâteaux differential, we treat $\mathbb{Q}$, and the continuous linear functional it defines as the same.

A functional $\rho:L^p\rightarrow\mathbb{R}\cup\{\infty\}$ is a risk measure, and it may possess the following properties:
	\begin{enumerate}
		\item Monotonicity: if $X \leq Y$, then $\rho(X) \geq \rho(Y),\:\forall\: X,Y\in L^p$.
		\item Translation Invariance: $\rho(X+c)=\rho(X)-c,\:\forall\: X\in L^p,\:\forall\:c \in \mathbb{R}$.
		\item Convexity: $\rho(\lambda X+(1-\lambda)Y)\leq \lambda \rho(X)+(1-\lambda)\rho(Y),\:\forall\: X,Y\in L^p,\:\forall\:\lambda\in[0,1]$.
  		\item Law Invariance: if $F_X=F_Y$, then $\rho(X)=\rho(Y),\:\forall\:X,Y\in L^p$.
    \item Comonotonic Additivity: $\rho(X+Y)=\rho(X)+\rho(Y)$ for any comonotonic pair $(X,Y)$.
	\end{enumerate}
	We have that  $\rho$ is called monetary if it fulfills (i) and (ii), convex if it is monetary and respects (iii), law invariant if it fulfills (iv), and comonotone if it has (v). We call a spectral risk measure if it possesses (i)-(v).  

  From Theorems 2.4, 2.11, and 3.1 of \cite{Kaina2009}, a map $\rho\colon L^p\rightarrow \mathbb{R}\cup\{\infty\}$, $p\in[1,\infty)$, is a convex risk measure if and only if it can be represented as
	\begin{equation*}\label{eq:dual}
	\rho(X)=\sup\limits_{\mathbb{Q}\in\mathcal{Q}}\left\lbrace E_\mathbb{Q}[-X]-\alpha_\rho(\mathbb{Q})\right\rbrace,\:\forall\:X\in L^p,
	\end{equation*} where  \begin{equation*}
\alpha_\rho(\mathbb{Q})=\sup\limits_{X\in L^p}\{E_\mathbb{Q}[-X]-\rho(X)\}.
	\end{equation*} Moreover, $\rho$ is finite if and only if the supremum over $\mathcal{Q}$ in its dual representation is attained. In this case, $\rho$ is continuous in the $L^p$ norm and in the bounded $\mathbb{P}$-a.s. convergence (Lebesgue continuous). For $p=\infty$, Theorem 4.33, Corollary 4.35 in \cite{Follmer2016} assures the claim under the additional assumption of Lebesgue continuity. In the finite case,  the maximum can be taken over a weakly compact $\mathcal{Q}^\prime\subseteq\mathcal{Q}$. We then assume, for the remainder of the paper, that this is the case.  Furthermore, since the maximum over $\mathcal{Q}^\prime$ in the dual representation is attained, Theorem 21 of \cite{Delbaen2012}, for $p=\infty$, and Theorem 3 of \cite{Ruszczynski2006}, for $p\in[1,\infty)$, assure that
       \[\partial \rho(X)=\left\lbrace \mathbb{Q}\in\mathcal{Q}\colon \rho(X)= E_\mathbb{Q}[-X]-\alpha_\rho(\mathbb{Q})\right\rbrace\neq\emptyset.\] Furthermore,
     $\rho$ is Gâteaux differentiable at $X$ if and only if $\partial \rho(X)=\{\mathbb{Q}\}$ is a singleton, which in this case the derivative turns out to be defined by $\mathbb{Q}$, i.e., the map $Z\mapsto E_\mathbb{Q}[-Z]$. 

     Under Law Invariance and $(\Omega,\mathcal{F},\mathbb{P})$ atomless, we have the special Kusuoka representation, see \cite{Filipovic2012} for $p\in[1,\infty)$ and \cite{Follmer2016} for $p=\infty$, \[\rho(X)=\max\limits_{\mathbb{Q}\in\mathcal{Q}}\left\lbrace\int_0^1F^{-1}_{-X}(u)F^{-1}_{\frac{d\mathbb{Q}}{d\mathbb{P}}}(u)du-\alpha_\rho(\mathbb{Q})\right\rbrace,\:\forall\:X\in L^p.\] In addition $\rho$ has Comonotonic Additivity, then we have the spectral representation as \[\rho_\phi(X)=-\int_0^1F^{-1}_X(u)\phi(u)du,\:\forall\:X \in L^p,\] where $\phi:[0,1]\to\mathbb{R}^+$ is a non-increasing functional such that $\int_{0}^{1}\phi(u)du=1$.

    \begin{Exm}\label{exm:rm}
    We now expose some concrete examples of law invariant convex risk measures we consider in this paper.
    \begin{enumerate}
        \item The Entropic risk measure (ENT) is the map $ENT_\gamma\colon L^\infty\to\mathbb{R}$  defined as \[ENT_\gamma (X) = \frac{1}{\gamma} \log \E [ e^{-\gamma X}],\:\gamma>0.\]  Its penalty is the relative entropy as \[\alpha_{ENT_\gamma}(\mathbb{Q})=\frac{1}{\gamma}E\left[\frac{d\mathbb{Q}}{d\mathbb{P}}\log\frac{d\mathbb{Q}}{d\mathbb{P}}\right].\] This risk measure is Gâteaux differentiable for any $X\in L^\infty$ with $\frac{d\mathbb{Q}_X}{d\mathbb{P}}=\frac{e^{-\gamma X}}{E[e^{-\gamma X}]}$. 
    \item The Shortfall Risk (SR) is a map defined as $SR_\ell\colon L^p\to\mathbb{R}$ as\[SR_\ell(X)=\inf\left\lbrace m\in\mathbb{R}\colon E[l(X-m)]\leq \ell_0\right\rbrace,\] where $l$  is a strictly convex and increasing loss function, and $\ell_0$ is an interior point in the range of $l$. A concrete and popular choice for loss function is the power functions given as $l(x)=\frac{1}{2}x^2 1_{x\geq 0}$. We then have that its penalty term is given, according to Example 4.118  of \cite{Follmer2016}, as \[\alpha_{SR_\ell}(\mathbb{Q})=(2\ell_0)^{\frac{1}{2}}\left\lVert \frac{d\mathbb{Q}}{d\mathbb{P}}\right\rVert_2.\] 
    \item Other example that is based on loss functions is the Optimized Certainty Equivalent (OCE), which is a map $OCE_\ell\colon L^p\to \mathbb{R}$ defined as \[OCE_\ell(X)=\inf\limits_{m\in\mathbb{R}}\left\lbrace E[l(X-m)]+m\right\rbrace,\] where $l$  is a strictly convex and increasing loss function. By Theorem 4.2 in \cite{Ben-Tal2007}, we have that $OCE_\ell$ is represented over $\alpha_{OCE_\ell}(\mathbb{Q})=E\left[l^*\left(\frac{d\mathbb{Q}}{d\mathbb{P}}\right)\right]$, where $l^*$ is the convex conjugate of $l$, sometimes called divergence between $\mathbb{Q}$ and $\mathbb{P}$.
    \item A example of a risk measure in $L^2$ is the Mean plus Semi-Deviation (MSD). Such risk measure is the functional $MSD_\beta: L^2 \rightarrow \mathbb{R}$ defined by\begin{align*} MSD_{\beta}(X)=-E[X] + \beta \lVert(X-E[X])^-\rVert_2, \beta\in[0,1].\end{align*} This risk measure is studied in detail by \cite{fischer2003risk}, belonging to the class of loss-deviation measures discussed by  \cite{righi2019composition}. It also possesses Positive Homogeneity as $\rho(\lambda X)=\lambda\rho(X)$ for any $\lambda\geq 0$ and any $X\in L^2$. In this case, Theorem 2.9 in \cite{Kaina2009} assures that a map $\rho\colon L^2\to\mathbb{R}$ is a convex risk measure with Positive Homogeneity, called coherent in the literature, if and only if it can be represented as 			\begin{equation*}\label{eq:cohdual} 			\rho(X)=\max\limits_{\mathbb{Q}\in\mathcal{Q}_\rho} E_\mathbb{Q}[-X],\:\forall\:X\in L^2, 			\end{equation*} where $\mathcal{Q}_\rho\subseteq \mathcal{Q}$ is a nonempty, closed, and convex set that is called the dual set of $\rho$. The dual set of MSD can be represented by \[\mathcal{Q}_{MSD_\beta}=\left \lbrace\mathbb{Q} \in \mathcal{Q} : \frac{d\mathbb{Q}}{d\mathbb{P}} =1+ \beta( V-E[V]), V\geq0, \lVert V\rVert_2=1 \right\rbrace.\] 
    \item The Expectile Value at Risk  (Exp), linked to the concept of an expectile. It is a functional $Exp_{\alpha}: L^1 \rightarrow \mathbb{R}$ directly defined as an argmin of a scoring function, which is given by	\begin{align*}	Exp_{\alpha}(X)&=-\argmin\limits_{x\in\mathbb{R}} E[\alpha[(X-x)^+]^2+(1-\alpha)[(X-x)^-]^2]=-e^\alpha(X),\:\alpha\in(0,1).	\end{align*} By \cite{Bellini2014}, the Exp is a law invariant coherent risk measure for $\alpha\leq0.5$. In addition, this measure is the only example of an elicitable coherent risk measure that does not collapse to the mean. See \cite{Ziegel2016} for details. The dual set of Exp can be given by \[\mathcal{Q}_{Exp_\alpha}=\left\lbrace\mathbb{Q}\in\mathcal{Q} \colon\:\exists\: a>0,\: a\leq\frac{d\mathbb{Q}}{d\mathbb{P}}\leq a\frac{1-\alpha}{\alpha} \right\rbrace.\]It is Gâteaux differentiable at any $X\in L^1$ with derivative $\mathbb{Q}_X$ defined as \[\frac{d\mathbb{Q}_X}{d\mathbb{P}}=\frac{\alpha 1_{X<e^\alpha(X)}+(1-\alpha)1_{X\geq e^\alpha(X)} }{E[\alpha 1_{X<e^\alpha(X)}+(1-\alpha)1_{X\geq e^\alpha(X)} ]}.\] 
    \item The most prominent example of spectral risk measure is the Expected Shortfall (ES), that is functional $ES_\alpha\colon L^1\to\mathbb{R}$ defined as \[ES_\alpha(X)=-\frac{1}{\alpha}\int_0^\alpha F^{-1}_X(u)du,\:\alpha\in(0,1).\]In this case the spectral function is $\phi(u)=\frac{1}{\alpha}1_{(0,\alpha)}(u),\:\alpha\in(0,1)$. The dual set of ES is defined as \[\mathcal{Q}_{ES_\alpha}=\left\lbrace\mathbb{Q}\in\mathcal{Q}\colon\frac{d\mathbb{Q}}{d\mathbb{P}}\leq\frac{1}{\alpha}\right\rbrace.\] It is Gateaux differentiable with  $\partial ES_\alpha(X)=\frac{1}{\alpha}1_{X\leq F^{-1}_X(\alpha)}$.
    \end{enumerate}
\end{Exm}

\section{General results}\label{proposed}

We now focus on exposing our proposed approach for robust convex risk measures. We begin with the formal definition of worst-case risk measure.

\begin{Def}
 Let $\rho$ be a law invariant convex risk measure. Its worst-case version under  a family of non-empty uncertainty sets $\{\mathcal{U}_X\subseteq L^p\}_{X\in L^p}$ is given as \begin{equation*}
    \rho^{WC}(X)=\sup\limits_{Z\in\mathcal{U}_X}\rho(Z).
\end{equation*}
\end{Def}

\begin{Rmk}
   There is preservation for the worst-case determination for operations preserved under point-wise supremum. More specifically, we have: if $\rho_1\geq \rho_2$, then $\rho_1^{WC}\geq\rho_2^{WC}$; $(\lambda\rho)^{WC}=\lambda\rho^{WC}$ for any $\lambda\geq 0$; $(\rho+c)^{WC}=\rho^{WC}-c$ for any $c\in\mathbb{R}$; and $(\sup_{i\in\mathcal{I}}\rho_i)^{WC}=\sup_{i\in\mathcal{I}}\rho_i^{WC}$, where $\mathcal{I}$ is arbitrary non-empty set.
\end{Rmk}

Next, to provide general results for worst-case law invariant convex risk measures, we utilize worst-case expectations as building blocks. We now have a simple but useful result that characterizes such maps as spectral risk measures.

\begin{Lmm}\label{Lmm:lema}
The family of maps \[f_Q\colon X\mapsto\sup_{X^\prime\sim X}E_Q[-X^\prime]=\int_0^1F^{-1}_{-X}(u)F^{-1}_{\frac{dQ}{d\mathbb{P}}}(u)du,\:Q\in\mathcal{Q},\] defines a spectral risk measure with $\phi_Q(u):=F^{-1}_{\frac{dQ}{d\mathbb{P}}}(1-u)$. In this case $\lVert\phi_Q\rVert_q=\left\lVert\frac{d Q}{d\mathbb{P}}\right\rVert_q$.
\end{Lmm}

\begin{proof}
It is an easy task to show that $\phi_Q(u):=F^{-1}_{\frac{dQ}{d\mathbb{P}}}(1-u)$ defines a valid distortion/spectral for any $Q\in\mathcal{Q}$. We also have $\lVert\phi_Q\rVert_q=\left\lVert\frac{d Q}{d\mathbb{P}}\right\rVert_q$. We show it for continuous $F_X$ by recalling that $U:=F_X(X)$ has uniform distribution over $(0,1)$. Nonetheless, the general case follows similar steps with more algebra under the modified distribution of $X$ given as $\Tilde{F}_X(x,\lambda)=\mathbb{P}(X<x)+\lambda\mathbb{P}(X=x)$, where $\lambda\in[0,1]$. In this case  if $\Tilde{U}$ is independent of $X$ and uniformly distributed over $(0,1)$, then we also have that $U:=\Tilde{F}_X(X,\Tilde{U})$ follows an uniform distribution over $(0,1)$.  We get that \begin{align*}\left\lVert\frac{dQ}{d\mathbb{P}} \right\rVert_q&=\left(\int_0^1\left(F^{-1}_{\frac{dQ}{d\mathbb{P}}}(1-u)\right)^qdu\right)^{\frac{1}{q}}=\left(\int_0^1(\phi_Q(u))^q du\right)^{\frac{1}{q}}=\lVert\phi_Q\rVert_q.\end{align*} 
\end{proof}

Given a specified radius $\epsilon>0$, closed balls in the canonical $L^p$ norm are $\mathcal{U}_X=X+\{Z\in L^p\colon\lVert Z\rVert_p\leq \epsilon \}$. For the Wasserstein distance or order $p\in[1,\infty)$, given as  \[d_{W_p}(X,Z)=\left(\int_0^1|F_X^{-1}(u)-F_Z^{-1}(u)|^pdu\right)^\frac{1}{p}\:p\in[1,\infty).\] For $p=\infty$ it is possible to defined $d_{W_\infty}(X,Z)=\lim_{p\to\infty}d_{W_p}(X,Z)$. For a detailed discussion on this metric, see \cite{Villani2021}, while \cite{Esfahani2018} is a reference for its use in robust decision-making. For some sensitivity analysis of risk measures in this context, see \cite{Bartl2021} and \cite{Nendel2022}.

In this case, $\mathcal{U}_{X,\epsilon}=\{Z\in L^p\colon  d_{W_p}(X,Z)\leq\epsilon\}$. We denote $\rho^{WC}_\epsilon$ the worst-case risk measure under this uncertainty set.
For Wasserstein closed balls and spectral risk measures, \cite{Liu2022} obtains the following formulation \[(\rho_{\phi})_\epsilon^{WC}(X)=\rho(X)+\epsilon\lVert \phi\rVert_q.\] We now extend such a result by showing that worst-case law invariant convex risk measures over closed balls in the $p$-norm and Wasserstein distance coincide. Further, it can be represented under a penalty term that is a scaled version of the original risk measure.

\begin{Thm}\label{thm:wasserstein}
\begin{align}
    \rho^{WC}_\epsilon(X)&=\max\limits_{\mathbb{Q}\in\mathcal{Q}}\left\lbrace E_\mathbb{Q}[-X]-\alpha_\rho(\mathbb{Q})+\epsilon\left\lVert \frac{d\mathbb{Q}}{d\mathbb{P}}\right\rVert_q\right\rbrace \label{eq:wass}\\
    &=\sup\{\rho(Z)\colon \lVert X-Z\rVert_p\leq \epsilon\},\:\forall\:X\in L^p.
\end{align}  The argmax  is 
\[X^*= \begin{cases}
    \left(X-\epsilon \left\lVert \frac{d\mathbb{Q}^*}{d\mathbb{P}}\right\rVert_\infty\right)1_A+X 1_{A^c},\:\mathbb{P}(A)=\left\lVert \frac{d\mathbb{Q}^*}{d\mathbb{P}}\right\rVert_\infty^{-1},& p=1,\\
   X-k\dfrac{d\mathbb{Q}^*}{d\mathbb{P}}^{\frac{q}{p}},& p\in(1,\infty),\\
   X-\epsilon, & p=\infty.
   \end{cases}\] where  $\mathbb{Q}^*$ is in the argmax of \eqref{eq:wass}, and $k$ solves  $d_{W_p}(X^*,X)=\epsilon$.
\end{Thm}

\begin{proof}
By \Cref{Lmm:lema}, we have $\sup\limits_{Z\in \mathcal{U}_{X,\epsilon}}f_Q(-Z)=f_Q(-X)+\epsilon\left\lVert \frac{dQ}{d\mathbb{P}}\right\rVert_q,\:\forall\:X\in L^p.$ Thus, we get  \begin{align*}
    \rho^{WC}_\epsilon(X)=\sup\limits_{d_{W_p}(X,Z)\leq \epsilon}\max\limits_{\mathbb{Q}\in\mathcal{Q}^\prime}\left\lbrace f_\mathbb{Q}(-Z)-\alpha_\rho(\mathbb{Q})\right\rbrace=\max\limits_{\mathbb{Q}\in\mathcal{Q}^\prime}\left\lbrace f_\mathbb{Q}(-X)+\epsilon\left\lVert \frac{dQ}{d\mathbb{P}}\right\rVert_q-\alpha_\rho(\mathbb{Q})\right\rbrace<\infty.
 \end{align*}
 Thus, $\rho^{WC}_\epsilon$ can be represented by $\mathbb{Q}\mapsto \alpha_\rho(\mathbb{Q})-\epsilon\left\lVert \frac{d\mathbb{Q}}{d\mathbb{P}}\right\rVert_q$. The last inequality follows since the maximum is attained in $\mathcal{Q}^\prime$ because $\mathbb{Q}\mapsto \epsilon\left\lVert \frac{dQ}{d\mathbb{P}}\right\rVert_q-\alpha_\rho(\mathbb{Q})$ is upper semicontinuous. From the penalty term obtained, we have that $\rho^{WC}_\epsilon$ is given as the sup-convolution; see \cite{Ekeland1999} or \cite{Zalinescu2002} for details, between  $\rho$ and the concave function defined as $X\mapsto -\mathbb{I}_{\lVert X\rVert_p\leq \epsilon}$. We then have for any $X\in L^p$ that \begin{align*}
        \rho^{WC}_\epsilon(X)&=\sup\limits_{Z\in L^p}\{\rho(X-Z)-\mathbb{I}_{\lVert Z\rVert_p\leq \epsilon}\}\\
        &=\sup\limits_{\lVert Z\rVert_p\leq \epsilon}\rho(X-Z)\\
        &=\sup\{\rho(Z)\colon \lVert X-Z\rVert_p\leq \epsilon\}.
    \end{align*}  

    Regarding the argmax, let $\mathbb{Q}^*$ be in the argmax of \eqref{eq:wass}. For $p=1$, let $X^*$ be such that \[\mathbb{P}\left(X^*=X-\epsilon \left\lVert \frac{d\mathbb{Q}^*}{d\mathbb{P}}\right\rVert_\infty\right)=\left(\left\lVert \frac{d\mathbb{Q}^*}{d\mathbb{P}}\right\rVert_\infty\right)^{-1}=1-\mathbb{P}(X^*=X).\] For $p\in(1,\infty)$, let $X^*=X-k\frac{d\mathbb{Q}^*}{d\mathbb{P}}^{\frac{q}{p}}$. Notice that $X^*\in L^p$. We can take $k$ such that $d_{W_p}(X^*,X)=\epsilon$. Then, we have that $X^*\in\mathcal{U}_{X,\epsilon}$. We also have that $|X-X^*|^p=k^p\frac{d\mathbb{Q}^*}{d\mathbb{P}}^q$. For $p=\infty$, let $X^*=X-\epsilon$. Recall that $\epsilon=d_{W_p}(X,X^*)\leq \lVert X-X^*\rVert_p$. Thus, for any $p\geq 1$ and  we have that \begin{align*}        \rho(X^*)&\geq E_{\mathbb{Q}^*}[X-X^*]+E_{\mathbb{Q}^*}[-X]-\alpha_\rho(\mathbb{Q}^*)\\                &=\lVert X-X^*\rVert_p \left\lVert \frac{d\mathbb{Q}^*}{d\mathbb{P}}\right\rVert_q+E_{\mathbb{Q}^*}[-X]-\alpha_\rho(\mathbb{Q}^*)\\        
   &\geq \epsilon \left\lVert \frac{d\mathbb{Q}^*}{d\mathbb{P}}\right\rVert_q+E_{\mathbb{Q}^*}[-X]-\alpha_\rho(\mathbb{Q}^*)= \rho^{WC}_\epsilon(X).    \end{align*} 
\end{proof}

\begin{Rmk} Let $C_X$ be the argmax of \eqref{eq:wass} for $X$. By \Cref{thm:wasserstein} we have that $C_X\cap\partial\rho(X)\neq\emptyset$ if and only if $\rho^{WC}_\epsilon(X)=\rho(X)+\epsilon\lVert\mathbb{Q}^*\rVert_q$ for any $\mathbb{Q}^*\in C_X\cap\partial\rho(X)\neq\emptyset$. This is the case for spectral risk measures, for instance. In this convenient setup, it is tempting to compute the sub-differential of $\rho$ as the Minkowski sum of the sub-differential of the two components. Nonetheless, since $X\mapsto\lVert \mathbb{Q}^*\rVert_q$, or even $X\mapsto\lVert \mathbb{Q}_X\rVert_q$ under Gâteaux differentiation, is in general not convex, we do not always have  $\partial\rho^{WC}_\epsilon(X)=\partial\rho(X)+\epsilon\partial\lVert \mathbb{Q}^*\rVert_q $. By recalling that the Fréchet derivative (or gradient) of the \( q \)-norm, $q\in (0,1)$, is given by \[\nabla \|\mathbb{Q}\|_q= \frac{|\mathbb{Q}|^{q-1} \text{sign}(\mathbb{Q})}{\|\mathbb{Q}\|_q^{q-1}},\] we have under $\rho$  be Gâteaux differentiable at $X$ and $h\colon X\mapsto\mathbb{Q}^*$ be Fréchet differentiable (or at least Gâteaux differentiable with continuous derivative) at $X$, that \[\partial\rho^{WC}_\epsilon(X)=\nabla\rho(X)+\epsilon\frac{{\mathbb{Q}^*}^{q-1}} {\|\mathbb{Q}^*\|_q^{q-1}}\nabla h(X)=\mathbb{Q}^*+\epsilon\frac{{\mathbb{Q}^*}^{q-1}} {\|\mathbb{Q}^*\|_q^{q-1}}\nabla h(X).\] 
\end{Rmk}

We now focus on the situation when the uncertainty set is determined by the moments (mean and variance) of the random variable. In this case we have that $\mathcal{U}_{X,\mu,\sigma}=\{Z\in L^2\colon E[Z]= E[X],\:\sigma(Z)=\sigma(X)\}$. For spectral risk measures
 results in \cite{Li2018}, \cite{Cornilly2018A}, \cite{Cornilly2019}, \cite{Cai2023}, \cite{Pesenti2020A}, \cite{Chen2021}, \cite{Shao2023}, \cite{Shao2023b},  \cite{Zuo2024} allow to conclude that  \[\rho_{\phi,\mu,\sigma}^{WC}(X)=-E[X]+\sigma(X)\lVert\phi-1\rVert_2,\] where the 2-norm is taken over $[0,1]$.  
 
 These authors also derive a closed form when $\rho$ is coherent and law invariant, relying on the fact that in this case $\rho=\sup_{\phi\in\Phi_\rho}\rho_\phi$, where $\Phi_\rho$. In this case the worst-case risk measure becomes  \[\rho^{WC}(X)=-E[X]+\sigma(X)\sup\limits_{\phi\in\Phi_\rho}\left\lVert\phi-1\right\rVert_2.\] Next, we expose a result that generalizes this framework by providing a formulation for worst-case law invariant convex risk measures over uncertainty sets based on mean and variance.

\begin{Thm}\label{prp:mean}
  \begin{equation}\label{eq:mean} \rho^{WC}_{\mu,\sigma}(X)=-E[X]+\max\limits_{\mathbb{Q}\in\mathcal{Q}}\left\lbrace\sigma(X)\left\lVert\frac{d\mathbb{Q}}{d\mathbb{P}}-1\right\rVert_2-\alpha_\rho(\mathbb{Q})\right\rbrace,\:\forall\:X\in L^2.\end{equation}  
The argmax is \[X^*=E[X]+\sigma(X)\dfrac{\left(\frac{d\mathbb{Q}^*}{d\mathbb{P}}-1\right)}{\left\lVert\frac{d\mathbb{Q}^*}{d\mathbb{P}}-1\right\rVert_2},\] 
where $\mathbb{Q}^*$ is in the argmax of \eqref{eq:mean}.
\end{Thm}

\begin{proof}

By  \Cref{Lmm:lema},  we get that \[\sup\limits_{Z\in \mathcal{U}_{X,\mu,\sigma}}f_Q(Z)=-E[X]+\sigma(X)\lVert \phi_Q-1\rVert_2,\:\forall\:X\in L^2,\:\forall\:Q\in\mathcal{Q}.\] We then get  for any $X\in L^2$ that \begin{align*}    \rho^{WC}_{\mu,\sigma}(X)&=E[X]+\max\limits_{\mathbb{Q}\in\mathcal{Q}}\left\lbrace\sigma(X)\left\lVert\frac{d\mathbb{Q}}{d\mathbb{P}}-1\right\rVert_2-\alpha_\rho(\mathbb{Q})\right\rbrace.\end{align*}
For the argmax,  if $X$ is constant, then the claim is trivial since $\mathcal{U}_{X,\mu,\sigma}=\{X\}$. Thus, fix a non-constant $X\in L^2$. It is straightforward to verify that $X^*\in\mathcal{U}_{X,\mu,\sigma}$. Thus,   $\rho(X^*)\leq\rho^{WC}_{\mu,\sigma}(X)$. On the other hand, we have that \begin{align*}\rho(X^*)&=-E[X]+\rho\left(\sigma(X)\left(\frac{d\mathbb{Q}^*}{d\mathbb{P}}-1\right)\left(\left\lVert\frac{d\mathbb{Q}^*}{d\mathbb{P}}-1\right\rVert_2\right)^{-1}\right)\\
=&-E[X]+\dfrac{\sigma(X)}{\left\lVert\frac{d\mathbb{Q}^*}{d\mathbb{P}}-1\right\rVert_2}\max\limits_{\mathbb{Q}\in\mathcal{Q}}\left\lbrace E\left[\left(\frac{d\mathbb{Q}^*}{d\mathbb{P}}-1\right)\left(\frac{d\mathbb{Q}}{d\mathbb{P}}-1\right)\right]-\dfrac{\left\lVert\frac{d\mathbb{Q}^*}{d\mathbb{P}}-1\right\rVert_2}{\sigma(X)}\alpha_\rho(\mathbb{Q}) \right\rbrace\\
\geq&-E[X]+\dfrac{\sigma(X)}{\left\lVert\frac{d\mathbb{Q}^*}{d\mathbb{P}}-1\right\rVert_2}\left[\left(\frac{d\mathbb{Q}^*}{d\mathbb{P}}-1\right)^{2}\right]-\dfrac{\left\lVert\frac{d\mathbb{Q}^*}{d\mathbb{P}}-1\right\rVert_2}{\sigma(X)}\alpha_\rho(\mathbb{Q}) \\
&=\rho^{WC}_{\mu,\sigma}(X).
\end{align*}
\end{proof}

\begin{Rmk}
If in addition to the conditions of \Cref{prp:mean}, $\rho$ fulfills Positive Homogeneity, then \[\rho^{WC}_{\mu,\sigma}(X)=-E[X]+\sigma(X)\max\limits_{\mathbb{Q}\in\mathcal{Q}_\rho}\left\lVert\frac{d\mathbb{Q}}{d\mathbb{P}}-1\right\rVert_2,\:\forall\:X\in L^2.\] The result for the closed form holds since $\alpha_\rho$ is the characteristic function of $\mathcal{Q}_\rho$. In this case,  $\rho^{WC}_{\mu,\sigma}$ is Gâteaux differentiable at any $X\in L^2$ with derivative \[\partial\rho^{WC}_{\mu,\sigma}(X)=1+\max\limits_{\mathbb{Q}\in\mathcal{Q}_\rho}\left\lVert\frac{d\mathbb{Q}}{d\mathbb{P}}-1\right\rVert_2\frac{X-E[X]}{\lVert X -E[X]\rVert_2}.\]Such claim follows by recalling that the expectation and the 2-norm are both Gâteaux differentiable with respective derivatives $1$ and $\frac{X}{\lVert X\rVert_2}$.
\end{Rmk}

\section{Concrete examples}\label{ConExamples}

We now focus on the concrete cases of the risk measures exposed in \Cref{exm:rm}. Our goal here is to apply the general results from the last section to obtain tractable closed forms. We begin with the Entropic. Recall that its domain is $L^\infty$. For the mean and variance uncertainty set, we have by the canonical extension of risk measures from $L^\infty$ to $L^2$, see \cite{Filipovic2012}, that if $X\in L^2$ with $Ent_\gamma(X)$ is finite, then $\partial Ent_\gamma(X)=\frac{e^{-\gamma X}}{E[e^{-\gamma X}]}\in L^2$.

\begin{Prp}\label{prp:ent}
\begin{enumerate}
    \item $(Ent_\gamma)^{WC}_\epsilon(X)=Ent_\gamma(X)+\epsilon$.
    \item $(Ent_\gamma)^{WC}_{\mu,\sigma}(X)=-E[X]+\sigma(X)\sqrt{e^{\gamma^2\sigma^2(X)}-1}-\frac{\gamma\sigma^2(X)}{2}$.
\end{enumerate}
\end{Prp}

\begin{proof}
For the uncertainty over Wasserstein closed ball, with $p=\infty$, \Cref{thm:wasserstein} indicates that the worst case risk measure simplifies to $(Ent_\gamma)_\epsilon^{WC}(X)=Ent_\gamma(X)+\epsilon$. For the uncertainty set over mean and variance, by \Cref{prp:mean}, we then seek to solve \[\max\limits_{\{Q\in L^2\colon Q\geq 0,E[Q]=1\}}\left\lbrace\sigma(X)\left\lVert Q-1\right\rVert_2-\frac{1}{\gamma}E[Q\log Q]\right\rbrace.\] The Lagrangian then becomes \[L(Q,\lambda)=\sigma(X)\left\lVert Q-1\right\rVert_2-\frac{1}{\gamma}E[Q\log Q]-\lambda(E[Q]-1),\lambda\leq 0.\] The first order condition then leads to
\[\frac{\sigma(X)(Q-1)}{\lVert Q-1\rVert_2}-\frac{1}{\gamma}(\log Q+1)=\lambda,\:\lambda\leq 0.\] By using the constraint $E[Q]=1$, we recover $\lambda =-\frac{1}{\gamma}E[(\log Q+1)]$. Thus, the first order condition implies \[\lVert \log Q -E[\log Q]\rVert_2=\sigma(\log Q)=\gamma\sigma(X).\] We can show that this is satisfied for a log-normal $Q^*=e^Y$, where $Y\sim N\left(-\frac{\gamma^2\sigma^2(X)}{2},\gamma^2\sigma^2(X)\right)$. In this case, we have that  \begin{align*}
    \max\limits_{\{Q\in L^2\colon Q\geq 0,E[Q]=1\}}\left\lbrace\sigma(X)\left\lVert Q-1\right\rVert_2-\frac{1}{\gamma}E[Q\log Q]\right\rbrace&=\sigma (X)\sigma(X)\sigma(Q^*)-\frac{1}{\gamma}E[Ye^Y]\\
    &=\sigma(X)\sqrt{e^{\gamma^2\sigma^2(X)}-1}-\frac{\gamma\sigma^2(X)}{2}.
\end{align*}
The last equality holds since $E[Ye^Y]=\frac{\partial}{\partial t}E[e^{tY}]|_{t=1}$ and noticing that $E[e^{tY}]=e^{\frac{(t^2-t)\gamma^2\sigma^2(X)}{2} }$. Hence, we obtain that \[(Ent_\gamma)^{WC}_{\mu,\sigma}(X)=-E[X]+\sigma(X)\sqrt{e^{\gamma^2\sigma^2(X)}-1}-\frac{\gamma\sigma^2(X)}{2}.\]
  \end{proof}

We now focus on the SR with the power loss function on $L^2$. Under certain parameter constraints, we obtain the following result.

\begin{Prp}\label{prp:sr}
\begin{enumerate}
    \item $(SR_\ell)_\epsilon^{WC}(X)
= \frac{(\epsilon-\sqrt{2\ell_0})^2 - E[X (X - \lambda)^+]}{E[(X - \lambda)^+]}$, where $\lambda$ solves $\|(X - \lambda)^+\|_2 = \epsilon-\sqrt{2\ell_0}$. If $(\epsilon-\sqrt{2\ell_0})^2-\sigma^2(X)>0$ and $\sqrt{(\epsilon-\sqrt{2\ell_0})^2-\sigma^2(X)}\geq E[X]-\essinf{X}$, then $(SR_\ell)_\epsilon^{WC}(X)=E[-X]+\sqrt{(\epsilon-\sqrt{2\ell_0})^2-\sigma^2(X)}$.
    \item  if $\sigma^2(X)>2\ell_0\geq \frac{\sigma^2(X)}{2}$, then $(SR_\ell)^{WC}_{\mu,\sigma}(X)=-E[X]+\frac{\sigma(X)\sqrt{4\ell_0-\sigma^2(X)}-2\ell_0}
{\sqrt{\sigma^2(X)-2\ell_0}}$.
\end{enumerate}
\end{Prp}

\begin{proof}
Regarding the uncertainty set based on Wasserstein closed balls, we have, by \Cref{thm:wasserstein}, to solve \[\max\limits_{\{Q\in L^2\colon Q\geq 0,E[Q]=1\}}\left\lbrace E[-XQ]+(\epsilon-\sqrt{2\ell_0})\left\lVert Q\right\rVert_2\right\rbrace.\] If $\epsilon=\sqrt{2\ell_0}$, then $SR_\ell^{WC}(X)=\infty$. Otherwise, the Lagrangian then becomes\[L(Q,\lambda)=E[-XQ]+(\epsilon-\sqrt{2\ell_0})\left\lVert Q\right\rVert_2-\lambda(E[Q]-1).\] The sub-differential condition   gives
\[
0 \leq -X + \lambda + (\epsilon-\sqrt{2\ell_0})\frac{Q}{\|Q\|_2}, \qquad Q \geq 0,\qquad Q \left(-X + \lambda + (\epsilon-\sqrt{2\ell_0})\frac{Q}{\|Q\|_2}\right) = 0.
\]
This yields the explicit solution
\[
Q^* = \frac{\|Q^*\|_2}{\epsilon-\sqrt{2\ell_0}}\left(X - \lambda\right)^+.\]
By imposing $E[Q^*]=1$, we get \[
\|Q^*\| = \frac{\epsilon-\sqrt{2\ell_0}}{E[(X - \lambda)^+]}.
\] Thus, the f.o.c. leads to 
\[
Q^* = \frac{(X - \lambda)^+}{E[(X - \lambda)^+]},\:\text{and}\:
\|(X - \lambda)^+\|_2 = \epsilon-\sqrt{2\ell_0}.
\] Hence, the maximum is 
\begin{align*}(SR_\ell)_\epsilon^{WC}(X)&=
E[-X Q^*] + (\epsilon-\sqrt{2\ell_0})\|Q^*\|_2 \\
&= -\frac{E[X (X - \lambda)^+]}{E[(X - \lambda)^+]} + \frac{(\epsilon-\sqrt{2\ell_0})^2}{E[(X - \lambda)^+]}\\
&= \frac{(\epsilon-\sqrt{2\ell_0})^2 - E[X (X - \lambda)^+]}{E[(X - \lambda)^+]},
\end{align*}
where $\lambda$ solves $\|(X - \lambda)^+\|_2 = \epsilon-\sqrt{2\ell_0}$.
In the special case of internal solution $Q>0$, the first order condition then becomes \[\frac{Q-1}{\lVert Q\rVert_2}=\frac{X-E[X]}{\epsilon-\sqrt{2\ell_0}}.\] In the case that $\epsilon-\sqrt{2\ell_0}>\sigma(X)$ and $\sqrt{(\epsilon-\sqrt{2\ell_0})^2-\sigma^2(X)}\geq E[X]-\essinf{X}$, we have that the argmax is given as \[\mathbb{Q}^*=1+\frac{X-E[X]}{\sqrt{(\epsilon-\sqrt{2\ell_0})^2-\sigma^2(X)}}.\] When this is the case, we have that \[(SR_\ell)_\epsilon^{WC}(X)=E\left[-XQ^*\right]+(\epsilon-\sqrt{2\ell_0})\left\lVert Q^*\right\rVert_2=E[-X]+\sqrt{(\epsilon-\sqrt{2\ell_0})^2-\sigma^2(X)}.\] 

For the uncertainty based on mean and variance, by \Cref{prp:mean}, in order to determine $SR_\ell^{WC}$ over uncertainty based on mean and variance we are interested in the value of \[\max\limits_{\mathbb{Q}\in\mathcal{Q}}\left\lbrace\sigma(X)\left\lVert\frac{d\mathbb{Q}}{d\mathbb{P}}-1\right\rVert_2-(2\ell_0)^{\frac{1}{2}}\left\lVert \frac{d\mathbb{Q}}{d\mathbb{P}}\right\rVert_2\right\rbrace.\] By recalling that$\left\lVert\frac{d\mathbb{Q}}{d\mathbb{P}}-1\right\rVert_2=\left(\left\lVert\frac{d\mathbb{Q}}{d\mathbb{P}}\right\rVert_2^2-1\right)^\frac{1}{2}$, we have that by making $y=E\left[\left(\frac{d\mathbb{Q}}{d\mathbb{P}}\right)^2\right]=\left\lVert\frac{d\mathbb{Q}}{d\mathbb{P}}\right\rVert_2^2$, the goal then becomes to determine the value of  \[\max\limits_{y\in [1,S]}\left\lbrace\sigma(X)(y-1)^\frac{1}{2}-(2\ell_0)^{\frac{1}{2}}y^{\frac{1}{2}}\right\rbrace,\] where $S$ is the $L^2$ bound of the weakly compact $\mathcal{Q}^\prime$. Thus, the critical point is obtained for $y=E\left[\left(\frac{d\mathbb{Q}}{d\mathbb{P}}\right)^2\right]=\frac{2\ell_0}{\sigma^2(X)-2\ell_0}$, which is valid when $\sigma^2(X)>2\ell_0\geq \frac{\sigma^2(X)}{2}$. Assuming this is the case, then we have that \[(SR_\ell)^{WC}_{\mu,\sigma}(X)=-E[X]+\frac{\sigma(X)\sqrt{4\ell_0-\sigma^2(X)}-2\ell_0}
{\sqrt{\sigma^2(X)-2\ell_0}}.\] 
\end{proof}

For the OCE, \cite{Bartl2020} obtain a robust formulation under Wasserstein closed balls as
\[(OCE_\ell)_\epsilon^{WC}(X)=\inf\limits_{\lambda\geq0}\left\lbrace OCE_{l_\lambda}(X)+\lambda\epsilon^p\right\rbrace,\] where $l_\lambda$ is a transform defined as \[l_\lambda(x)=\sup\limits_{l(y)<\infty}\{l(y)-\lambda |x-y|^p\}.\]  We now show that this coincides with the result obtained by our approach. 

\begin{Prp}\label{prp:OCE}
\[(OCE_\ell)_\epsilon^{WC}(X)=\inf\limits_{\lambda\geq 0}\left\lbrace OCE_{l_\lambda}(X)+\lambda\epsilon\right\rbrace=\sup\limits_{\mathbb{Q}\in\mathcal{Q}}\left\lbrace E_\mathbb{Q}[-X]-E\left[l^*\left(\frac{d\mathbb{Q}}{d\mathbb{P}}\right)\right]+\epsilon\left\lVert\frac{d\mathbb{Q}}{d\mathbb{P}}\right\rVert_q\right\rbrace.\]
\end{Prp}

\begin{proof}
We show for OCE with $p\in(1,\infty)$. The claims for $p=1$ and $p=\infty$ are simpler.  Further, for each $\lambda\geq 0$, we have by calculation that $(l_\lambda)^*(y)=l^{*}(y)-(\lambda p)^{1-q}q^{-1}|y|^q$. We then obtain the following:
\begin{align*}
\inf\limits_{\lambda\geq0}\left\lbrace OCE_{l_\lambda}(X)+\lambda\epsilon\right\rbrace&=\inf\limits_{\lambda\geq 0}\sup\limits_{\mathbb{Q}\in\mathcal{Q}}\left\lbrace E_\mathbb{Q}[-X]-E\left[l^*\left(\frac{d\mathbb{Q}}{d\mathbb{P}}\right)\right]+\frac{\left\lVert\frac{d\mathbb{Q}}{d\mathbb{P}}\right\rVert^q_q}{(\lambda p)^{q-1}q}+\epsilon^p\lambda\right\rbrace.
\end{align*}
However, for any $\mathbb{Q}\in\mathcal{Q}$ the convex map \[h(\lambda)=\frac{\left\lVert\frac{d\mathbb{Q}}{d\mathbb{P}}\right\rVert^q_q}{(\lambda p)^{q-1}q}+\epsilon^p\lambda\] attains its infimum, which is, obtained  after a very tedious calculation, $\epsilon \left\lVert\frac{d\mathbb{Q}}{d\mathbb{P}}\right\rVert_q$. Hence, we recover our result since 
\[\inf\limits_{\lambda\geq 0}\left\lbrace OCE_{l_\lambda}(X)+\lambda\epsilon\right\rbrace=\sup\limits_{\mathbb{Q}\in\mathcal{Q}}\left\lbrace E_\mathbb{Q}[-X]-E\left[l^*\left(\frac{d\mathbb{Q}}{d\mathbb{P}}\right)\right]+\epsilon\left\lVert\frac{d\mathbb{Q}}{d\mathbb{P}}\right\rVert_q\right\rbrace.\]
\end{proof}

The next Proposition deals with the concrete example of MSD. Recall that its domain is the $L^2$.

\begin{Prp}\label{prp:MSD}
    \begin{enumerate}
        \item $(MSD_\beta)^{WC}_\epsilon(X)=MSD_\beta(X)+\epsilon\sqrt{1+\beta^2}$.
        \item $(MSD_\beta)^{WC}_{\mu,\sigma}(X)=-E[X]+\beta\sigma(X)$.
    \end{enumerate}
\end{Prp}

\begin{proof}

By \Cref{thm:wasserstein} we have that under uncertainty over the Wasserstein closed ball \[(MSD_\beta)_\epsilon^{WC}(X)=\max\limits_{\mathbb{Q}\in\mathcal{Q}_\rho}\left\lbrace E_\mathbb{Q}[-X]+\epsilon\left\lVert\frac{d\mathbb{Q}}{d\mathbb{P}}\right\rVert_2 \right\rbrace.\] For any $\mathbb{Q}\in\mathcal{Q}_{MSD_\beta}$ we have that \[\left\lVert \frac{d\mathbb{Q}}{d\mathbb{P}}\right\rVert_2=\lVert 1+\beta(V-E[V])\rVert_2=\sqrt{1+\beta^2}.\] We then have \[(MSD_\beta)_\epsilon^{WC}(X)=\max\limits_{\mathbb{Q}\in\mathcal{Q}_\rho}\left\lbrace E_\mathbb{Q}[-X]+\epsilon\sqrt{1+\beta^2} \right\rbrace=MSD_\beta(X)+\epsilon\sqrt{1+\beta^2}.\]
For the mean and variance based uncertainty, \Cref{prp:mean} implies that for any coherent risk measure we have \[\rho^{WC}_{\mu,\sigma}(X)=-E[X]+\sigma(X)\max\limits_{\mathbb{Q}\in\mathcal{Q}_\rho}\left\lVert\frac{d\mathbb{Q}}{d\mathbb{P}}-1\right\rVert_2,\:\forall\:X\in L^2.\]  Notice that for any $\mathbb{Q}\in\mathcal{Q}_{MSD_\beta}$ we have that \[\left\lVert \frac{d\mathbb{Q}}{d\mathbb{P}}-1\right\rVert_2=\beta\lVert V-E[V]\rVert_2=\beta\sqrt{E[V^2]-E[V]^2}.\] Since $V\geq0$ and $E[V^2]=1$, we have that $\lVert V-E[V]\rVert_2\leq 1$. By taking $V =\frac{(X-E[X])^-}{\lVert(X-E[X])^-\rVert_2}$ for $X\in L^2$, we have that $\lVert V-E[V]\rVert_2=1$. Thus, in light of \Cref{prp:mean}, we have that  \begin{align*}    (MSD_\beta)^{WC}_{\mu,\sigma}(X)&=-E[X]+\sigma(X)\max\limits_{\mathbb{Q}\in\mathcal{Q}_{MSD_\beta}}\left\lVert \frac{d\mathbb{Q}}{d\mathbb{P}}-1\right\rVert_2=-E[X]+\beta\sigma(X),\:\forall\:X\in L^2. \end{align*}
\end{proof}

For the Expectile, under Wasserstein closed balls based uncertainty, there is, in general, no closed-form solution for its worst-case representation (see Theorem 2 in \cite{Hu2024}, for instance).  For mean and variance-based uncertainty, \cite{Hu2024} computes the worst-case scenario by considering a representation of expectiles as the point-wise maximum of spectral risk measures. We now show that this coincides with our result. 

\begin{Prp}\label{prp:exp}
    $(Exp_\alpha)^{WC}_{\mu,\sigma}(X)=-E[X]+\sigma(X)\frac{(\beta-1)\sqrt{\beta}}{2\beta}$, where $\beta=\frac{1-\alpha}{\alpha}$.
\end{Prp}

\begin{proof}
In order to obtain $(Exp_\alpha)^{WC}_{\mu,\sigma}$, we must compute $\max_{\mathbb{Q}\in\mathcal{Q}_{Exp_\alpha}}\left\lVert\frac{d\mathbb{Q}}{d\mathbb{P}}-1\right\rVert_2$. Due to the nature of $\mathcal{Q}_{Exp_\alpha}$, this is a tricky quest. Nonetheless, in Proposition 9 of \cite{Bellini2014} a formulation for Exp is given as \[Exp^{\alpha}(X)=\max\limits_{\gamma\in\left[\frac{1}{\beta},1\right]}\left\lbrace(1-\gamma)ES_{\tau}(X)+\gamma E[-X] \right\rbrace,\:\tau=1-\frac{\beta-\frac{1}{\gamma}}{\beta-1}.\] According to Theorem 7 of \cite{Hu2024}, this maximum is attained for $\gamma^*=\frac{\frac{1}{\beta}+1}{2}$. This representation leads to \[\mathcal{Q}_{Exp_\alpha}=\left(1-\frac{\frac{1}{\beta}+1}{2}\right)\mathcal{Q}_{ES_\alpha}+\frac{\frac{1}{\beta}+1}{2}\mathbb{P}.\] From \Cref{prp:mean}, we must to compute \[\max\limits_{\mathbb{Q}\in\mathcal{Q}_{Exp_\alpha}}\left\lVert\frac{d\mathbb{Q}}{d\mathbb{P}}-1\right\rVert_2=\max\limits_{\{\mathbb{Q}\in\mathcal{Q}\colon \frac{d\mathbb{Q}}{d\mathbb{P}}\leq\frac{1}{\tau*}\}}\left\lVert\left(1-\frac{\frac{1}{\beta}+1}{2}\right)\frac{d\mathbb{Q}}{d\mathbb{P}}+\frac{\frac{1}{\beta}+1}{2}-1\right\rVert_2.\] Since $\left\lVert\frac{d\mathbb{Q}}{d\mathbb{P}}-1\right\rVert_2=E\left[\frac{d\mathbb{Q}}{d\mathbb{P}}^2\right]-1$, we have that the maximum over $\{\mathbb{Q}\in\mathcal{Q}\colon \frac{d\mathbb{Q}}{d\mathbb{P}}\leq\frac{1}{\tau*}\}$ is attained by the same extreme point $\mathbb{Q}^*$ with derivative \[\frac{d\mathbb{Q}^*}{d\mathbb{P}}=\left(1-\frac{\frac{1}{\beta}+1}{2}\right)\frac{1}{\tau*}1_{U\leq\tau^*}+\frac{\frac{1}{\beta}+1}{2},\] where $U$ is an uniform random variable over $[0,1]$. Direct calculation then gives us that $\left\lVert\frac{d\mathbb{Q}^*}{d\mathbb{P}}-1\right\rVert_2=\frac{(\beta-1)\sqrt{\beta}}{2\beta}$. Hence, we have that \[(Exp_\alpha)^{WC}(X)=-E[X]+\sigma(X)\frac{(\beta-1)\sqrt{\beta}}{2\beta}.\]
\end{proof}

For the ES, since it is comonotone additive, we recover the known result from the literature. We now demonstrate that this result aligns with the one predicted by our proposed approach.

\begin{Prp}\label{prp:ES}
    \begin{enumerate}
        \item \[(ES_\alpha)^{WC}_\epsilon(X)=\begin{cases}ES_\alpha(X)+\frac{1}{\alpha}\epsilon,&\:p=1\\
    ES_\alpha(X)+\left(\frac{1}{\alpha}\right)^{\frac{1}{q}}\epsilon,&\:p\in(1,\infty)\\
    ES_\alpha(X)+\epsilon,&\:p=\infty.\end{cases}.\]
        \item $(ES_\alpha)^{WC}_{\mu,\sigma}(X)=-E[X]+\sigma(X)\sqrt{\frac{1-\alpha}{\alpha}}$.
    \end{enumerate}
\end{Prp}

\begin{proof}
For the uncertainty regarding the Wasserstein closed balls we have to maximize  $\frac{1}{\alpha}1_{X\leq F^{-1}_X(\alpha)}+\epsilon\left\lVert\frac{d\mathbb{Q}}{d\mathbb{P}}\right\rVert_q$ over $\mathcal{Q}_{ES_\alpha}=\left\lbrace\mathbb{Q}\in\mathcal{Q}\colon\frac{d\mathbb{Q}}{d\mathbb{P}}\leq\frac{1}{\alpha}\right\rbrace$. It is straightforward to observe that, for each $X\in L^p$, the Gateaux derivative $\partial ES_\alpha(X)=\frac{1}{\alpha}1_{X\leq F^{-1}_X(\alpha)}$ is the argmax of both $\mathbb{Q}\mapsto\frac{1}{\alpha}1_{X\leq F^{-1}_X(\alpha)}$, by the definition of sub-differential, and $\mathbb{Q}\mapsto\left\lVert\frac{d\mathbb{Q}}{d\mathbb{P}}\right\rVert_q$, by definition of $\mathcal{Q}_{ES_\alpha}$. Thus, a direct calculation delivers the result.

For the uncertainty based on mean and variance, it is clear that $\sqrt{\frac{1-\alpha}{\alpha}}=\sqrt{\frac{1}{\alpha}-1}\geq\max\limits_{\mathbb{Q}\in\mathcal{Q}_\rho}\left\lVert\frac{d\mathbb{Q}}{d\mathbb{P}}-1\right\rVert_2$. For the converse inequality, for each $X\in L^2$, we have that $\frac{\mathbb{Q}_X}{d\mathbb{P}}=\frac{1}{\alpha}1_{X\leq F^{-1}_X(\alpha)}\in\mathcal{Q}_{ES^\alpha}$. Then, we obtain that \[\max\limits_{\mathbb{Q}\in\mathcal{Q}_\rho}\left\lVert\frac{d\mathbb{Q}}{d\mathbb{P}}-1\right\rVert_2\geq\sup_{X\in L^2}\left\lVert \frac{1}{\alpha}1_{X\leq F^{-1}_X(\alpha)}-1\right\rVert_2=\sqrt{\frac{1-\alpha}{\alpha}}.\]
\end{proof}

\section{Capital determination}\label{Sub1}

To assess the practical implications of the theoretical contributions developed in this paper, we conduct numerical simulations of a capital determination problem under different uncertainty sets (e.g., $p$-norm/Wasserstein distance and moment-based).
We consider that the financial position $X$ is defined in a discrete probability space $\Omega = (w_1, \cdots, w_T)$, as $X(w_t) = X_t$, for $t = 1, \cdots, T$, where $T$ represents the number of observations. Thus, we have $\mathbb{P}(X = X_t) = \mathbb{P}(w_t) = \frac{1}{T}$, which results in the empirical distribution and expectation given by $F_X(x) = \frac{1}{T}\sum_{t=1}^T 1_{X_t \leq x}$ and $E[X] = \frac{1}{T}\sum_{t=1}^T X_t$, respectively. Based on the empirical distribution, we employ historical simulation (HS) as a method for risk estimation. This approach is non-parametric and does not rely on any specific assumptions regarding the distribution of the data, making it particularly suitable for capturing empirical characteristics of financial returns. Furthermore, it serves as a foundational approach for risk estimation \citep{kuester2006value, Righi2020}, and has also been adapted for use in robust risk estimation \citep{Bartl2021, bernard2024robust}. In this case, $\frac{d\mathbb{Q}^*}{d\mathbb{P}}(w_t) = T\mathbb{Q}^*(w_t)$, for $t = 1, \cdots, T$, and any probability measure $\mathbb{Q}^*$ is absolutely continuous with respect to $\mathbb{P}$.

Table~\ref{cenario} summarizes the distributions used in the simulations, which cover stylized features commonly observed in financial data: normality, skewness, and heavy tails.
We fix the sample size at $T = 1{,}000$ in all simulations, in line with practices commonly adopted in the literature on robust risk measures (e.g., \cite{zhu2009worst} and \cite{bernard2024robust}).
Estimation of worst-case measures is based on two sets of uncertainties that are aligned with our theoretical framework. 
For each set, we estimate capital requirements using the risk measures defined in Section \ref{ConExamples}.
We report results for MSD (with $\beta=1$) and ES (with $\alpha=2.5\%$ and $5\%$), as both admit closed-form worst-case solutions (Propositions~\ref{prp:MSD} and~\ref{prp:ES}) and are commonly used in regulatory contexts.
After estimating the risk, we use the results for capital determination. As the monetary value of the investment, we assume a value equal to 1. This assumption simplifies the interpretation of the results by aligning capital requirements with the numerical value of the risk measure, that is, 
\textit{CP := Capital Determination = Worst-case risk measure value × 1}.  For each scenario, we repeat the CP estimation 10,000 times using Monte Carlo methods to ensure the robustness and replicability of the results. Therefore, in our description, we present the average values.

\begin{center}
    $<$ Insert Table \ref{cenario}. $>$
\end{center}

\begin{figure}[ht!]
\centering
\includegraphics[width=1\linewidth]{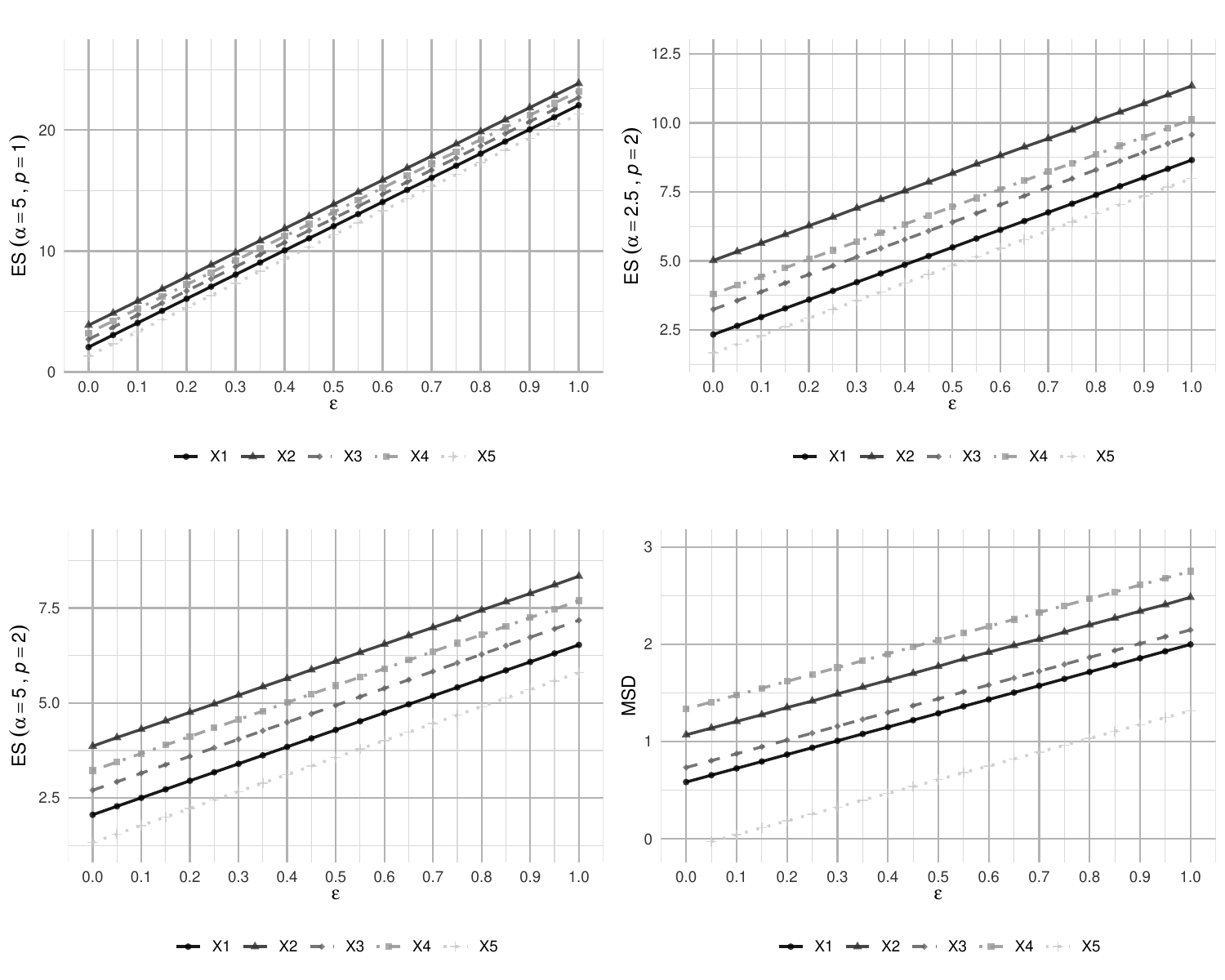}\\
\caption{Capital determination derived from ES (with $\alpha = 2.5\%$ and $5\%$, and $p = 2$) and MSD, considering uncertainty sets based on closed balls under $p$-norms and the Wasserstein distance for varying values of $\epsilon$.}\label{fig1}
\tiny \singlespace \justifying {{Note: The variables follow the probability specifications and parameters described in Table \ref{cenario}.}  \par}
\end{figure}

Figure \ref{fig1} illustrates the results of capital determination,  considering both uncertainty sets for $X_1$ to $X_5$. For worst-case measures of uncertainty sets based on closed balls under the $p$-norms and the Wasserstein
distance, we varied the parameter $\epsilon$ $0$ to $1$ in increments of $0.05$, allowing us to examine how increasing levels of uncertainty impact capital determination. A similar choice for $\epsilon$ was made by \cite{jin2024constructing}. 
For $(ES_\alpha)^{WC}_\epsilon$, we consider $p=1$ and $2$. It can be observed that capital determination increases as the uncertainty parameter $\epsilon$ increases. This finding is consistent for all risk measures and financial positions. As the uncertainty parameter \( \epsilon \) increases, the uncertainty set \( \mathcal{U}_X \) expands, and the worst-case risk measure rises accordingly. This effect occurs because the term $\epsilon\left\lVert \frac{d\mathbb{Q}}{d\mathbb{P}} \right\rVert$  is explicitly introduced as a linear penalization component,  proportionally amplifying capital values across all scenarios while preserving their relative rankings.

\begin{center}
    $<$ Insert Figure \ref{fig1}.  $>$
\end{center}

It is noted that the worst-case measures maintain patterns consistent with those of the baseline measures (without uncertainty) concerning the relative behavior among the distributions and risk parameters. 
As expected, distributions with heavier tails (e.g., \(X_2\)) or negative skewness (e.g., \(X_4\)) yield higher capital requirements, while symmetric or positively skewed distributions (e.g., \(X_1\), \(X_5\)) show lower values.
The capital determination is higher to compensate for the increased risk associated with scenarios of more severe or frequent losses. Furthermore, ES with $\alpha=2.5\%$ is systematically greater than with $\alpha=5\%$, reflecting the greater severity associated with a more extreme quantile of the loss distribution. This consistent pattern is observed in both the baseline and worst-case measures (with $\epsilon>0$), as the proportional effect of $\epsilon$ amplifies the absolute values while maintaining the hierarchy between scenarios and parameters.

The analysis of $(ES_\alpha)^{WC}_\epsilon$ for $p=1$ and $p=2$ reveals that the norm parameter used to define the uncertainty set has a direct impact on the quantification of the worst-case measure. Consistent with Proposition~\ref{prp:ES}, the choice of the norm parameter \( p \) affects the degree of penalization: \( p = 1 \) results in systematically higher capital requirements than \( p = 2 \), providing stronger protection against uncertainty in the financial variable.


\begin{center}
    $<$ Insert Table \ref{result1}.$>$
\end{center}

Table \ref{result1} presents the capital requirements obtained using $(ES_{\alpha})^{WC}_{\mu,\sigma}$ and $(MSD_{\beta})^{WC}_{\mu,\sigma}$ for positions $X_1$ to $X_8$. It also includes the non-robust (baseline) estimates for comparison.
 $(ES_{\alpha})^{WC}_{\mu,\sigma}$ and $(MSD_{\beta})^{WC}_{\mu,\sigma}$ lead to more conservative and robust capital determinations. 
In absolute terms, $X_6$ and $X_7$ exhibit the highest values for $(ES_{\alpha})^{WC}_{\mu,\sigma}$. 
Both financial positions have a higher variance ($\sigma^2 = 2$), leading to a stronger penalization when using an uncertainty set defined by mean and variance. As shown in Proposition~\ref{prp:ES}, the robust measure based on the moment $(ES^{WC}_{\mu,\sigma})$ is given by the mean plus a penalty term proportional to the standard deviation, scaled by $\sqrt{(1 - \alpha)/\alpha}$. Thus, distributions with greater dispersion result in higher capital requirements, reflecting the increased sensitivity of the robust measure to volatility under this specification.
On the other hand, $X_5$ and $X_8$ display the lowest absolute values for both $ES_{\alpha}$ and $ES^{WC}_{\mu,\sigma}$, but show the largest relative increases compared to their baseline measures. This occurs because the uncertainty set considers distributions with the same mean and standard deviation as the original variable, yet allows for completely different distributional shapes, particularly ones that may place more weight on adverse tail outcomes. Therefore, the robust measure is capable of capturing hidden or underestimated risks that the baseline measure overlooks.

\begin{figure}[H]
    \centering
    \includegraphics[width=0.95\textwidth]{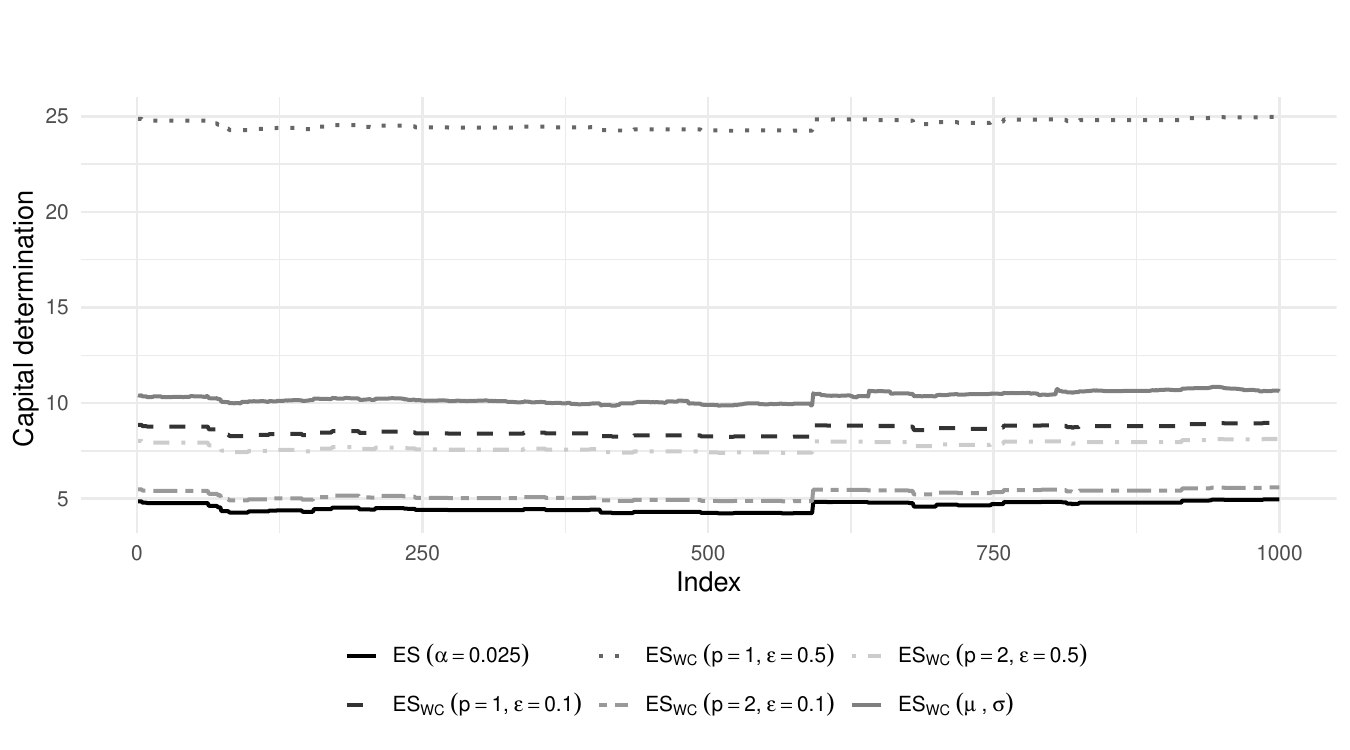}
    \caption{Capital Determination for $X_4$ considering $ES_{2.5\%}$, $(ES_{2.5\%})^{WC}_\epsilon$ ($\epsilon = 0.1$ and 0.5, $p = 1$ and 2) $(ES_{2.5\%})^{WC}_{\mu,\sigma}$.}
    \label{fig:determinacao-capital}
    \tiny \singlespace \justifying {{Note: In the figure, ES ($\alpha = 0.025$) represents the baseline risk measure; ES$_{WC}$ $(p=1, \epsilon=0.1)$ corresponds to $(ES_{2.5\%})^{WC}_{0.1}$ with $p=1$; ES$_{WC}$ $(p=1, \epsilon=0.5)$ corresponds to $(ES_{2.5\%})^{WC}_{0.5}$ with $p=1$; ES$_{WC}$ $(p=2, \epsilon=0.1)$ corresponds to $(ES_{2.5\%})^{WC}_{0.1}$ with $p=2$; ES$_{WC}$ $(p=2, \epsilon=0.5)$ corresponds to $(ES_{2.5\%})^{WC}_{0.5}$ with $p=2$; and ES$_{WC}(\mu, \sigma)$ denotes $(ES_{2.5\%})^{WC}_{\mu,\sigma}$.}  \par}
\end{figure}

In Figure \ref{fig:determinacao-capital}, we present an illustration of capital determination for a sample of 2,000 observations from $X_4$, using a rolling estimation window of 1,000 observations. Capital requirements were computed using $ES_{2.5\%}$, the $(ES_\alpha)^{WC}_\epsilon$ with $\epsilon = 0.1$ and $0.5$ for $p = 1$ and $2$, and the  $(ES_{2.5\%})^{WC}_{\mu,\sigma}$. As expected, higher values of $\epsilon$ and lower values of $p$ (e.g., $p=1$) lead to more conservative estimates, due to broader uncertainty sets and stronger penalization. Moment-based uncertainty, while preserving the mean and variance, allows for heavier tails and more adverse scenarios, especially under high volatility. 
Our results show that capital determination is shaped by how uncertainty about the random variable \( X \) is modeled. Worst-case convex risk measures capture the consequences of such uncertainty by enlarging the risk evaluation domain beyond the observed data. In this sense, capital requirements rise not only due to tail behavior, but because the true distribution of \( X \) may deviate from its empirical counterpart.

\begin{center}
    $<$ Insert Figure \ref{fig:determinacao-capital} $>$
\end{center}



\section{Portfolio Application}\label{Sub2}

In this section, we illustrate the application of worst-case convex risk measures in a portfolio optimization setting. 
We consider an arbitrage-free market represented by a set of non-redundant payoffs $Y=\{Y_1,\dots,Y_d\}\in(L^\infty)^d$. Non-redundancy is linear independence in that $\sum_{i=1}^dw_iY_i=0$ if and only if $w_i=0$ for any $i=1,\dots,d$. This assumption ensures the uniqueness of the representation in the optimization process.

Given this market structure, the set of admissible portfolio weights is defined as \[\mathcal{X}=\left\lbrace X=(w_1,\dots,w_d)\in[0,1]^d\colon\sum_{i=1}^dw_i=1,\sum_{i=1}^dw_i \mathbb{E}[Y_i]\geq 0\right\rbrace.\]
The goal is to identify the composition $X\in\mathcal{X}$ that minimizes the risk of the portfolio.  
The corresponding optimization problem is given by
    \[\begin{aligned}	P(\rho):=&\inf\limits_{X\in\mathcal{X}}\rho\left(\sum_{i=1}^dw_iY_i\right)\\		&{\rm s.t.}\\		&w_i\geq0,\:\forall\:i=1,\dots,d\\		&\sum_{i=1}^dw_i=1.\\		
    \end{aligned}\]	
This formulation excludes short selling and requires full capital allocation, 
constraints that are standard in portfolio optimization models \citep{Righi2018com}.

For the risk measure $\rho$, our focus is on $(SR_\ell)_\epsilon^{WC}$, $(SR_\ell)_{\mu,\sigma}^{WC}$ and $(Ent_\gamma)^{WC}_{\mu,\sigma}$. This choice is motivated by one of the main contributions of our study, which is to present closed-form solutions for these measures under the specific uncertainty sets explored in this analysis. 
Proposition~\ref{Prop7} provides explicit formulations for portfolio optimization problems under $(SR_\ell)_\epsilon^{WC}$, $(SR_\ell)_{\mu,\sigma}^{WC}$ and $(Ent_\gamma)^{WC}_{\mu,\sigma}$. For $(SR_\ell)_\epsilon^{WC}$, we present two versions based on \Cref{prp:sr}.

  \begin{Prp}\label{Prop7}
Let $\mu_Y=(E[Y_1],\dots, E[Y_d])$ and $\Sigma_Y$ the covariance matrix of $Y$. Then
\begin{enumerate}
 \item \[\begin{aligned}
P\left((SR_\ell)_\epsilon^{WC}\right)&:=\inf\limits_{X\in\mathcal{X},\:\lambda\in\mathbb{R},\:G\in\mathbb{R}^T}\frac{(\epsilon-\sqrt{2\ell_0})^2-\frac{1}{T}\sum_{i=1}^T(X^\prime Y)G}{\frac{1}{T}\sum_{i=1}^TG}\\
    s.t.&\\
    &G\geq X^\prime Y\\
    &G\geq0\\
    &\frac{1}{T}\sum_{i=1}^TG^2=(\epsilon-\sqrt{2\ell_0})^2.
    \end{aligned}\]
    \item \[\begin{aligned}
P^\prime\left((SR_\ell)_\epsilon^{WC}\right)&:=\inf\limits_{X\in\mathcal{X}}-X^\prime\mu_Y+\sqrt{(\epsilon-\sqrt{2\ell_0})^2-X^\prime\Sigma_Y X}\\
    s.t.&\\
    &(\epsilon-\sqrt{2\ell_0})^2-X^\prime\Sigma_Y X\geq0\\
    &-X^\prime\mu_Y+\sqrt{(\epsilon-\sqrt{2\ell_0})^2-X^\prime\Sigma_Y X}\geq \min X^\prime Y.
    \end{aligned}\]
    \item \[\begin{aligned}
P\left((SR_\ell)_{\mu,\sigma}^{WC}\right)&:=\inf\limits_{X\in\mathcal{X}}-X^\prime\mu_Y+\dfrac{\sqrt{X^\prime\Sigma_Y X(4\ell_0-X^\prime\Sigma_Y X)}-2\ell_0}{\sqrt{X^\prime\Sigma_Y X-2\ell_0}}\\
    s.t.&\\
    &X^\prime\Sigma_Y X>2\ell_0\\
    &2\ell_0\geq \frac{X^\prime\Sigma_Y X}{2}.
    \end{aligned}\]
    \item \[P\left((Ent_\gamma)^{WC}_{\mu,\sigma}\right):=\inf\limits_{X\in\mathcal{X}}-X^\prime\mu_Y+\sqrt{X^\prime\Sigma_Y X}\sqrt{e^{\gamma^2X^\prime\Sigma_Y X}-1}-\frac{\gamma X^\prime\Sigma_Y X}{2}.\]
\end{enumerate}
  \end{Prp}

\begin{proof}
Direct application from the results in \Cref{ConExamples}.  
\end{proof}

We estimate the risk measures using Historical Simulation. Regarding the parameters, we set $\epsilon = 0.1, 0.5, \text{and}\, 1.0$ for $(SR_\ell)_\epsilon^{WC}$, and $\lambda = 1$ for $(Ent_\gamma)^{WC}_{\mu,\sigma}$.  
Our portfolios are constructed with 4 and 16 assets. Concerning the data-generating process, we consider three data-generating processes. The first consists of samples drawn from a multivariate normal distribution with a mean of zero and a variance of one. The second scenario assumes a multivariate $t$-Student distribution with three degrees of freedom. The third scenario is a mixture model composed of assets following a normal or a $t$-Student distribution.  
Additionally, we account for both low and high correlation settings (0.2 and 0.8). As a result, we analyze a total of six distinct scenarios. These scenarios are inspired by the study of \cite{muller2018numerical}.

Table~\ref{Tab:port} presents the descriptive analysis of the portfolios composed of four assets under different distribution and correlation scenarios.
The results for the 16-asset portfolios are available upon request.
We report the average value of the weights $w_i,\ i = 1, \cdots, 4$, as well as the concentration of each portfolio, calculated as
$\text{Conc} = \sum_{i=1}^d w_i^2.$
The \( \text{Conc} \) metric varies only due to the Monte Carlo procedure for portfolios without replication; higher values of \( \text{Conc} \) indicate a higher concentration of the portfolio.
In addition, we calculate the average return (Average), the standard deviation (SD), and the VaR at the significance level 5\% (VaR$_{5\%}$) based on the portfolio return distribution.
Since a total of 10,000 portfolios were generated, our final results represent the average value of each descriptive measure.

\begin{center}
    $<$ Insert Table \ref{Tab:port}. $>$
\end{center}

The results of the Entropic risk measure provide evidence that the portfolios obtained using the robust approach $(\text{Ent}_\gamma)^{WC}_{\mu,\sigma}$ are more diversified compared to the non-robust version $(\text{Ent}_\gamma)$. The weights assigned to the assets under $(\text{Ent}_\gamma)^{WC}_{\mu,\sigma}$ are more homogeneous, especially in scenarios with normally distributed assets and low correlation, where the weights are nearly uniform.
In contrast, for scenarios generated with mixture distributions or the $t$-Student distribution, the non-robust version $(\text{Ent}_\gamma)$ results in more concentrated allocations. 
This pattern in the weight statistics is reflected in the portfolio concentration metric (Conc), which is lower for $(\text{Ent}_\gamma)^{WC}_{\mu,\sigma}$ compared to $(\text{Ent}_\gamma)$ in the same scenario. For illustration, consider the \textit{$t$ 0.80} scenario, where the robust risk measure yields a $Con$ of 0.3494, whereas the non-robust version reaches 0.6540.
This corresponds to a reduction of approximately 46.6\% in portfolio concentration, indicating an improvement in diversification.
We also observe differences in the risk and return statistics of the portfolios.
Although $(\text{Ent}_\gamma)^{WC}_{\mu,\sigma}$ leads to a reduction in the average return - for instance, from 0.0072 to 0.0018 in the $t$ 0.20 scenario - is accompanied by a decrease in standard deviation (SD from 1.1453 to 1.0567) and tail risk (VaR$_{5\%}$ from 1.6333 to 1.4970). 
This reduction, compared to $(\text{Ent}_\gamma)$, tends to be more pronounced in scenarios involving assets with a high tail, which may indicate better protection against losses and volatility under such conditions.
When evaluating portfolios composed of 16 assets, similar results are observed. However, it should be noted that in these portfolios, $(\text{Ent}_\gamma)^{WC}_{\mu,\sigma}$ leads to a more pronounced reduction in concentration under adverse scenarios, particularly those characterized by high correlation and heavy-tailed distributions.

The comparison between the results of the Shortfall Risk measure in its baseline form $(SR_\ell)$ and its robust counterpart under moment-based uncertainty $(SR_\ell)^{WC}_{\mu,\sigma}$ reveals behavioral patterns similar to those observed for the Entropic risk measure.
The robust version tends to produce less concentrated portfolios, mainly in scenarios with a mixture of distributions, including those with low and high correlations, where the robust measure achieves better weight diversification. With regard to risk and return statistics, it is observed that the robust measure, especially in mixture scenarios, leads to reductions in both volatility and tail risk. For example, in the Mixture 0.8 scenario, SD drops from 1.7342 to 1.1867 and VaR$_{5\%}$ from 2.3445 to 1.6456, while the average return remains relatively stable (0.0043 to 0.0025).

The analysis of the results for the measure \((SR_\ell)^{WC}_{\epsilon}\) indicates that the concentration of optimized portfolios tends to remain stable as the values of \(\epsilon\) vary. Compared to the baseline version, the robust measure leads to a lower concentration - and therefore a greater diversification - especially for \(\epsilon = 0.5\), except in the \textit{Normal 0.20} scenario, where this pattern does not hold. In terms of performance metrics, the results are generally similar to those of the baseline. For \(\epsilon = 0.5\), the SD and \(\text{VaR}_{5\%}\) show moderate reductions (e.g., SD from 1.7869 to 1.7572 in $t$ 0.20),  indicating improved portfolio stability and reduced tail risk under this level of robustness.
When applied to portfolios with 16 assets, $(SR_\ell)^{WC}_{\epsilon}$ generally enhances risk-adjusted performance, particularly for $\epsilon = 0.5$.  
In scenarios characterized by high correlation and heavy tails — such as those generated from $t$-Student or Mixture distributions — this formulation tends to reduce portfolio concentration, thereby promoting greater diversification.  This improved diversification is achieved without significant loss in average return, suggesting that moderate levels of robustness ($\epsilon = 0.5$) strike a favorable balance between risk mitigation and profitability.

When comparing the results obtained from $(SR_\ell)^{WC}_{\mu,\sigma}$ and $(SR_\ell)^{WC}_{\epsilon}$, differences are observed in both portfolio concentration and volatility. In most scenarios, the moment-based approach leads to lower concentration and reduced SD. For example, in the $t$ 0.20 scenario, the concentration under $(SR_\ell)^{WC}_{\mu,\sigma}$ is 0.4693, compared to 0.9692 under $(SR_\ell)^{WC}_{0.5}$—an approximate reduction of 106.5\%. In the same setting, the standard deviation decreases from 1.7572 to 1.3546, representing a reduction of about 22.9\%.
Similarly, in the Mixture 0.2 scenario, concentration drops from 0.9734 to 0.3711 (a 162.3\% reduction), while the SD declines from 1.7221 to 0.9604—an improvement of approximately 44.2\%. These gains indicate greater diversification and lower risk under the robust formulation based on moment constraints.
This pattern is consistently observed across the analyzed scenarios. A thorough comparison of portfolio performance across both uncertainty sets remains beyond the scope of this section.



\bibliography{Theory,reference}
\bibliographystyle{apalike}





\newpage

\begin{table}[ht!]
\centering
\caption{Specification of probability distributions and parameters for the numerical example of capital determination}
\label{cenario}
\begin{tabular}{lccc}
\hline
\textbf{Financial position} & \textbf{Distribution} & \textbf{Abb.} & \textbf{Parameters} \\
\hline
$X_1$ & Normal & N & $\mu = 0, \sigma^2 = 1$\\
$X_2$ &Student's $t$ & $t$ & $\nu = 3$\\
$X_3$ &Student's $t$ & $t$ & $\nu = 6$\\
$X_4$ &Skewed $t$ & AST &   $\lambda = -1$, $\nu = 6$ \\
$X_5$ &Skewed $t$ & AST & $\lambda = 1$, $\nu = 6$ \\
$X_6$ & Normal & N & $\mu = 0, \sigma^2 = 2$\\
$X_7$ & Normal & N & $\mu = 1, \sigma^2 = 2$\\
$X_8$ & Normal & N & $\mu = 1, \sigma^2 = 1$\\
\hline 
\end{tabular}
{\tiny \singlespace \justifying {{Note: Abb. means abbreviation. The parameters $\mu$, $\sigma^2$, $\nu$, and $\lambda$ represent the mean, variance, degrees of freedom, and skewness, respectively. $X_1$ follows a standard normal distribution, with mean zero and variance one. $X_2$ and $X_3$ have Student's $t$ distribution.  $X_2$, with $\nu = 3$, has a higher probability of extreme losses and a highly volatile time series. $X_3$, with $\nu = 6$, exhibits more moderately heavy tails, approximating the empirical behavior of stock indices in moderate market conditions. For example, when analyzing the historical time series of the S\&P 500 index from April 29, 2021, to April 25, 2025, we observe an empirical kurtosis value similar to that obtained from the simulated series.
We also include skewed $t$ distributions. The process $X_4$, with a skewness parameter of $\lambda = -1$, represents a left-skewed distribution. In contrast, $X_5$, with $\lambda = 1$, indicates a distribution distorted to the right. $X_6, X_7, \text{and}\; X_8$ are normally distributed but with means or variances different from those of $X_1$.}  \par}}
\end{table}

\begin{table}[ht!]
\centering
\caption{Capital determination was obtained using ES at significance levels of $\alpha=2.5\%$ and $\alpha=5\%$, as well as using MSD, under an uncertainty set defined by the mean and variance. Additionally, capital determination was derived based on the baseline ES and MSD.}
\label{result1}
\begin{tabular}{lcccccc}
\hline
\textbf{Financial position} & $(ES_{2.5\%})^{WC}_{\mu,\sigma}$ & $(ES_{5\%})^{WC}_{\mu,\sigma}$ & \textbf{$(MSD_\beta)^{WC}_{\mu,\sigma}$} & $ES_{2.5\%}$ & $ES_{5\%}$ & {$MSD_\beta$}\\
\hline
$X_1$ & 6.2435 & 4.3579 & 0.9998 & 2.3292 & 2.0580 & 0.5835 \\
$X_2$ & 10.6626 & 7.4424 & 1.7078 & 5.0111 & 3.8606 & 1.0665 \\
$X_3$ & 7.6466 & 5.3371 & 1.2241 & 3.2422 & 2.7036 & 0.7328 \\
$X_4$ & 7.1307 & 5.1731 & 1.6870 & 3.7974 & 3.2230 & 1.3354 \\
$X_5$ & 5.8281 & 3.8718 & 0.3879 & 1.6588 & 1.3268 & -0.0983 \\
$X_6$ & 12.4888 & 8.7172 & 2.0004 & 4.6604 & 4.1184 & 1.1680 \\
$X_7$ & 11.4922 & 7.7197 & 1.0013 & 3.6630 & 3.1195 & 0.1685 \\
$X_8$ & 5.2438 & 3.3580 & -0.0004 & 1.3296 & 1.0579 & -0.4168 \\
\hline 
\end{tabular}
\tiny \singlespace \justifying {{Note: The variables follow the probability specifications and parameters detailed Table \ref{cenario}.}  \par}
\end{table}

\begin{table}[ht]
\centering
\footnotesize 
\caption{Statistical summary of portfolios with four assets in six scenarios.}
\renewcommand{\arraystretch}{1.1}  
\setlength{\tabcolsep}{3pt}  
\begin{tabular}{l@{\hskip 5pt}r@{\hskip 5pt}r@{\hskip 5pt}r@{\hskip 5pt}r@{\hskip 5pt}r@{\hskip 5pt}r@{\hskip 5pt}r@{\hskip 5pt}r}
  \hline
 \textbf{Scenarios} & $w_1$ & $w_2$ & $w_3$ & $w_4$ & Con & Average & SD & VaR$_{5\%}$ \\ 
 \hline
 \multicolumn{9}{c}{\textbf{Baseline measures}}\\
  \hline
   \multicolumn{9}{c}{$(Ent_\gamma)$} \\
  \hline
  Normal 0.20 & 0.2497 & 0.2496 & 0.2504 & 0.2502 & 0.2556 & 0.0029 & 0.6339 & 1.0355 \\ 
  Normal 0.80 & 0.2487 & 0.2500 & 0.2515 & 0.2499 & 0.2855 & 0.0031 & 0.9223 & 1.5083 \\ 
  $t$ 0.20 & 0.2506 & 0.2484 & 0.2523 & 0.2486 & 0.3659 & 0.0072 & 1.1453 & 1.6333 \\ 
  $t$ 0.80 & 0.2492 & 0.2496 & 0.2531 & 0.2482 & 0.6540 & 0.0060 & 1.6129 & 2.2581 \\ 
  Mixture 0.2 & 0.3799 & 0.3795 & 0.1204 & 0.1203 & 0.3240 & 0.0034 & 0.6709 & 1.0788 \\ 
  Mixture 0.8 & 0.3840 & 0.3851 & 0.1155 & 0.1154 & 0.3437 & 0.0036 & 0.8231 & 1.3267 \\ 
  \hline
 \multicolumn{9}{c}{$(SR_\ell)$} \\
  \hline
Normal 0.20 & 0.2550 & 0.2492 & 0.2514 & 0.2444 & 0.9905 & -0.0003 & 1.0042 & 1.6390 \\ 
  Normal 0.80 & 0.2446 & 0.2509 & 0.2474 & 0.2571 & 0.9967 & -0.0000 & 1.0063 & 1.6402 \\ 
 $t$ 0.20 & 0.2551 & 0.2558 & 0.2456 & 0.2436 & 0.9967 & 0.0172 & 1.7869 & 2.3343 \\ 
   $t$ 0.80 & 0.2472 & 0.2450 & 0.2549 & 0.2529 & 0.9739 & 0.0102 & 1.7613 & 2.3368 \\ 
  Mixture 0.2 & 0.0010 & 0.0011 & 0.5020 & 0.4959 & 0.9931 & 0.0073 & 1.7389 & 2.3375 \\ 
  Mixture 0.8 & 0.0001 & 0.0001 & 0.4952 & 0.5046 & 0.9951 & 0.0043 & 1.7342 & 2.3445 \\ 
  \hline
 \multicolumn{9}{c}{\textbf{Robust risk measures}} \\
  \hline
 \multicolumn{9}{c}{$(Ent_\gamma)^{WC}_{\mu,\sigma}$} \\
  \hline
Normal 0.20 & 0.2497 & 0.2497 & 0.2503 & 0.2503 & 0.2533 & 0.0020 & 0.6324 & 1.0348 \\ 
Normal 0.80 & 0.2494 & 0.2497 & 0.2513 & 0.2496 & 0.2665 & 0.0015 & 0.9203 & 1.5080 \\ 
$t$ 0.20 & 0.2496 & 0.2502 & 0.2503 & 0.2499 & 0.2661 & 0.0018 & 1.0567 & 1.4970 \\ 
$t$ 0.80 & 0.2495 & 0.2514 & 0.2510 & 0.2481 & 0.3494 & 0.0014 & 1.5482 & 2.1748 \\ 
Mixture 0.2 & 0.3687 & 0.3687 & 0.1314 & 0.1312 & 0.3098 & 0.0018 & 0.6646 & 1.0675 \\ 
Mixture 0.8 & 0.3687 & 0.3694 & 0.1312 & 0.1307 & 0.3173 & 0.0016 & 0.8140 & 1.3085 \\ 
\hline
  \multicolumn{9}{c}{ $(SR_\ell)_{\mu,\sigma}^{WC}$} \\
\hline
Normal 0.20 & 0.2495 & 0.2469 & 0.2560 & 0.2477 & 0.5163 & 0.0139 & 0.7926 & 1.2863 \\ 
  Normal 0.80 & 0.2501 & 0.2511 & 0.2492 & 0.2497 & 0.9965 & 0.0107 & 1.0083 & 1.6435 \\ 
  $t$ 0.20 & 0.2482 & 0.2533 & 0.2494 & 0.2490 & 0.4693 & 0.0123 & 1.3546 & 1.7723 \\ 
  $t$ 0.80 & 0.2511 & 0.2489 & 0.2495 & 0.2504 & 0.9915 & 0.0096 & 1.7898 & 2.3448 \\ 
  Mixture 0.2 & 0.1941 & 0.1939 & 0.3096 & 0.3025 & 0.3711 & 0.0049 & 0.9604 & 1.3567 \\ 
  Mixture 0.8 & 0.1498 & 0.1504 & 0.3496 & 0.3502 & 0.3626 & 0.0025 & 1.1867 & 1.6456 \\ 
  \hline
  \multicolumn{9}{c}{$(SR_\ell)_{0.1}^{WC}$} \\
  \hline
Normal 0.20 & 0.2561 & 0.2498 & 0.2509 & 0.2431 & 0.9909 & -0.0003 & 1.0044 & 1.6393 \\ 
  Normal 0.80 & 0.2468 & 0.2503 & 0.2457 & 0.2573 & 0.9986 & 0.0001 & 1.0058 & 1.6391 \\ 
 $t$ 0.20 & 0.2560 & 0.2557 & 0.2423 & 0.2459 & 0.9922 & 0.0174 & 1.7820 & 2.3302 \\ 
  $t$ 0.80 & 0.2479 & 0.2462 & 0.2543 & 0.2516 & 0.9821 & 0.0095 & 1.7618 & 2.3391 \\ 
  Mixture 0.2 & 0.0009 & 0.0009 & 0.4998 & 0.4984 & 0.9936 & 0.0072 & 1.7398 & 2.3378 \\ 
  Mixture 0.8 & 0.0001 & 0.0001 & 0.4930 & 0.5067 & 0.9941 & 0.0043 & 1.7337 & 2.3441 \\ 
    \hline
     \multicolumn{9}{c}{ $(SR_\ell)_{0.5}^{WC}$} \\
  \hline
Normal 0.20 & 0.2528 & 0.2470 & 0.2492 & 0.2510 & 0.9941 & 0.0035 & 1.0038 & 1.6354 \\ 
  Normal 0.80 & 0.2459 & 0.2537 & 0.2483 & 0.2521 & 0.9578 & 0.0027 & 1.0007 & 1.6317 \\ 
  t 0.20 & 0.2529 & 0.2523 & 0.2472 & 0.2476 & 0.9692 & 0.0124 & 1.7572 & 2.3082 \\ 
  t 0.80 & 0.2480 & 0.2479 & 0.2547 & 0.2494 & 0.9738 & 0.0069 & 1.7563 & 2.3370 \\ 
  Mixture 0.2 & 0.0001 & 0.0001 & 0.5035 & 0.4964 & 0.9734 & 0.0063 & 1.7221 & 2.3172 \\ 
  Mixture 0.8 & 0.0003 & 0.0002 & 0.4957 & 0.5038 & 0.9823 & 0.0031 & 1.7270 & 2.3416 \\ 
\hline
  \multicolumn{9}{c}{ $(SR_\ell)_{1}^{WC}$} \\
  \hline
Normal 0.20 & 0.2552 & 0.2491 & 0.2497 & 0.2460 & 0.9973 & -0.0003 & 1.0069 & 1.6432 \\ 
  Normal 0.80 & 0.2463 & 0.2507 & 0.2473 & 0.2557 & 0.9749 & 0.0029 & 1.0035 & 1.6348 \\ 
  t 0.20 & 0.2544 & 0.2552 & 0.2437 & 0.2467 & 0.9800 & 0.0173 & 1.7713 & 2.3161 \\ 
  t 0.80 & 0.2482 & 0.2474 & 0.2551 & 0.2493 & 0.9877 & 0.0082 & 1.7612 & 2.3403 \\ 
  Mixture 0.2 & 0.0000 & 0.0001 & 0.5043 & 0.4956 & 0.9928 & 0.0068 & 1.7404 & 2.3382 \\ 
  Mixture 0.8 & 0.0001 & 0.0002 & 0.4946 & 0.5051 & 0.9918 & 0.0044 & 1.7332 & 2.3438 \\ 
  \hline
\end{tabular}\label{Tab:port}
  \tiny \singlespace \justifying {{Note: In the figure, $w_i,\ i = 1, \cdots, 4$, refers to the weight of each asset. Con stands for portfolio concentration. Average represents the {mean portfolio return}, SD is the standard deviation, while $\text{VaR}_{5\%}$ denotes the 5\% Value at Risk of the portfolio returns. In the scenarios labeled {0.20} and {0.80}, the multivariate distribution was generated assuming asset correlations of 0.2 and 0.8, respectively. The {Normal} scenario refers to a multivariate normal distribution with zero mean and unit variance. The \textit{t} scenario corresponds to a multivariate $t$-Student distribution with 3 degrees of freedom. In the {Mixture} scenario, assets 1 and 2 follow a normal distribution, while assets 3 and 4 follow a $t$-Student distribution.}  \par}
\end{table}

\end{document}